\documentclass[letterpaper, amsfonts, superscriptaddress, amssymb, amsmath, preprint, showkeys, nofootinbib, raggedbottom]{revtex4-2}

\usepackage{mathptmx}
\usepackage[utf8]{inputenc}
\usepackage{amsmath}
\usepackage{xcolor}
\usepackage{graphicx}
\usepackage{adjustbox}
\usepackage[T1]{fontenc}
\usepackage[left]{lineno}
\usepackage{array}

\newcommand{\angstrom}{\mbox{\normalfont\AA}}

\usepackage[pdftex, pdftitle={Article}, pdfauthor={Author}]{hyperref} 
\graphicspath{ {Figures/} }

\begin{document}
\title{Electrical switching of an unconventional odd-parity magnet}

\affiliation{Department of Physics, Massachusetts Institute of Technology, Cambridge, MA 02139, USA.}
\affiliation{Department of Materials Science and Engineering, Massachusetts Institute of Technology, Cambridge, MA 02139, USA.}
\affiliation{Consiglio Nazionale delle Ricerche CNR-SPIN, c/o Universit\`{a} degli Studi `G. D'Annunzio', Chieti, 66100, Italy}
\affiliation{Vin\v{c}a Institute of Nuclear Sciences - 
National Institute of the Republic of Serbia, University of Belgrade, P. O. Box 522, RS-11001 Belgrade, Serbia}
\affiliation{Consiglio Nazionale delle Ricerche CNR-SPIN, Area della Ricerca di Tor Vergata, Via del Fosso del Cavaliere 100, I-00133 Rome, Italy}
\affiliation{Department of Molecular Sciences and Nanosystems, Ca’ Foscari University of Venice, 30123 Venice, Italy}
\affiliation{Department of Physics, Yale University, New Haven, CT 06511, USA.}
\affiliation{Department of Physics, Drexel University, Philadelphia, PA 19104, USA.}
\affiliation{Rutgers Center for Emergent Materials and Department of Physics and Astronomy, Rutgers University, NJ 08854, USA.}
\affiliation{Department of Physics, The Grainger College of Engineering, University of Illinois Urbana-Champaign, Urbana, IL 61801, USA.}
\affiliation{Anthony J. Leggett Institute for Condensed Matter Theory, The Grainger College of Engineering, University of Illinois Urbana-Champaign, Urbana, 61801, IL, USA.}
\affiliation{Department of Materials Science, University of Milan-Bicocca, 20125 Milan, Italy}

\author{Qian Song}
\email{qiansong@mit.edu}
\affiliation{Department of Physics, Massachusetts Institute of Technology, Cambridge, MA 02139, USA.}
\affiliation{Department of Materials Science and Engineering, Massachusetts Institute of Technology, Cambridge, MA 02139, USA.}

\author{Srdjan Stavri\'{c}}
\affiliation{Consiglio Nazionale delle Ricerche CNR-SPIN, c/o Universit\`{a} degli Studi `G. D'Annunzio', Chieti, 66100, Italy}
\affiliation{Vin\v{c}a Institute of Nuclear Sciences - 
National Institute of the Republic of Serbia, University of Belgrade, P. O. Box 522, RS-11001 Belgrade, Serbia}

\author{Paolo Barone}
\affiliation{Consiglio Nazionale delle Ricerche CNR-SPIN, Area della Ricerca di Tor Vergata, Via del Fosso del Cavaliere 100, I-00133 Rome, Italy}

\author{Andrea Droghetti}
\affiliation{Consiglio Nazionale delle Ricerche CNR-SPIN, c/o Universit\`{a} degli Studi `G. D'Annunzio', Chieti, 66100, Italy}
\affiliation{Department of Molecular Sciences and Nanosystems, Ca’ Foscari University of Venice, 30123 Venice, Italy}

\author{Daniil S. Antonenko}
\affiliation{Department of Physics, Yale University, New Haven, CT 06511, USA.}

\author{J\"{o}rn W. F. Venderbos}
\affiliation{Department of Physics, Drexel University, Philadelphia, PA 19104, USA.}

\author{Connor A. Occhialini}
\affiliation{Department of Physics, Massachusetts Institute of Technology, Cambridge, MA 02139, USA.}

\author{Batyr Ilyas}
\affiliation{Department of Physics, Massachusetts Institute of Technology, Cambridge, MA 02139, USA.}

\author{Emre Ergeçen}
\affiliation{Department of Physics, Massachusetts Institute of Technology, Cambridge, MA 02139, USA.}

\author{Nuh Gedik}
\affiliation{Department of Physics, Massachusetts Institute of Technology, Cambridge, MA 02139, USA.}

\author{Sang-Wook Cheong}
\affiliation{Rutgers Center for Emergent Materials and Department of Physics and Astronomy, Rutgers University, NJ 08854, USA.}

\author{Rafael M. Fernandes}
\email{rafaelf@illinois.edu}
\affiliation{Department of Physics, The Grainger College of Engineering, University of Illinois Urbana-Champaign, Urbana, IL 61801, USA.}
\affiliation{Anthony J. Leggett Institute for Condensed Matter Theory, The Grainger College of Engineering, University of Illinois Urbana-Champaign, Urbana, 61801, IL, USA.}

\author{Silvia Picozzi}
\affiliation{Consiglio Nazionale delle Ricerche CNR-SPIN, c/o Universit\`{a} degli Studi `G. D'Annunzio', Chieti, 66100, Italy}
\affiliation{Department of Materials Science, University of Milan-Bicocca, 20125 Milan, Italy}

\author{Riccardo Comin}
\email{rcomin@mit.edu}
\affiliation{Department of Physics, Massachusetts Institute of Technology, Cambridge, MA 02139, USA.}


\maketitle

\newpage
\textbf{Abstract}

\textbf{Magnetic states with zero magnetization but non-relativistic spin splitting are outstanding candidates for the next generation of spintronic devices. Their electron-volt (eV) scale spin splitting, ultrafast spin dynamics and nearly vanishing stray fields make them particularly promising for several applications \cite{bai2023efficient, baltz2018antiferromagnetic, olejnik2018terahertz}. A variety of such magnetic states with nontrivial spin textures have been identified recently, including even-parity \textit{d}, \textit{g}, or \textit{i}-wave altermagnets and odd-parity \textit{p}-wave magnets \cite{vsmejkal2022emerging, vsmejkal2022beyond, Hellenes2024pwavemagnet, krempasky2024altermagnetic, fedchenko2024observation}. Achieving voltage-based control of the nonuniform spin polarization of these magnetic states is of great interest for realizing energy-efficient and compact devices for information storage and processing \cite{wadley2016electrical, yan2020electric}. Spin-spiral type-II multiferroics are optimal candidates for such voltage-based control, as they exhibit an inversion-symmetry-breaking magnetic order which directly induces ferroelectric polarization, allowing for symmetry protected cross-control between spin chirality and polar order \cite{tokura2014multiferroics, fiebig2016evolution, Kurumaji2020, masuda2021electric, cheong2007multiferroics}. Here we combine photocurrent measurements, first-principle calculations, and group-theory analysis to provide direct evidence that the spin polarization of the spin-spiral type-II multiferroic NiI$_2$ exhibits odd-parity character connected to the spiral chirality. The symmetry-protected coupling between chirality and polar order enables electrical control of a primarily non-relativistic spin polarization. Our findings represent the first direct observation of unconventional odd-parity magnetism in a spin-spiral type-II multiferroic, and open a new frontier of voltage-based switching of non-relativistic spin polarization in compensated magnets.}

\section*{Introduction}


Recent advances in spin-group classification \cite{vsmejkal2022beyond,vsmejkal2022emerging, chen2025} have led to the prediction and subsequent observation of collinear altermagnets \cite{krempasky2024altermagnetic, fedchenko2024observation}, a new class of magnets with fully compensated magnetic moments but exhibiting strong non-relativistic spin splitting, i.e., spin splitting that persists even without spin-orbit coupling (SOC) \cite{Hayami2019,yuan2020giant, yuan2021prediction}. While non-collinear magnetic states can also exhibit spin splitting \cite{chen2023octupole, qin2023room, zhu2024observation, cheong2024altermagnetism}, recent spin-group analysis has revealed that certain non-collinear compensated magnetic configurations can sustain sizable odd-parity spin polarization even in the absence of SOC \cite{Hellenes2024pwavemagnet, brekke2024minimal} (see Extended Data Figure 1). Here, spin polarization refers to the expectation value of the spin angular momentum operator over Bloch states in momentum space. A key feature of odd-parity magnets is that the inversion operation reverses the spin polarization, suggesting the possibility of controlling the primarily non-relativistic spin texture electrically when SOC is accounted for.


Non-collinear spin spiral (type-II) multiferroics exhibit inversion-symmetry-breaking magnetic order characterized by spin chirality (the handedness of the helix), which can induce improper ferroelectric polarization via SOC\cite{katsura2005spin, sergienko2006role}, allowing for electrical control of compensated magnetic states \cite{babkevich2012electric, stein2017control, sagayama2010observation, yamasaki2007electric}. While previous studies have focused on the interplay between spin chirality and ferroelectricity in spin spiral multiferroics, here we demonstrate that predominantly non-relativistic spin polarization is also present and can be controlled electrically.

\section{Symmetries of the spin helix in NiI$_2$}

\begin{figure*}[htb!]

\centering
\includegraphics[width=0.95\textwidth]{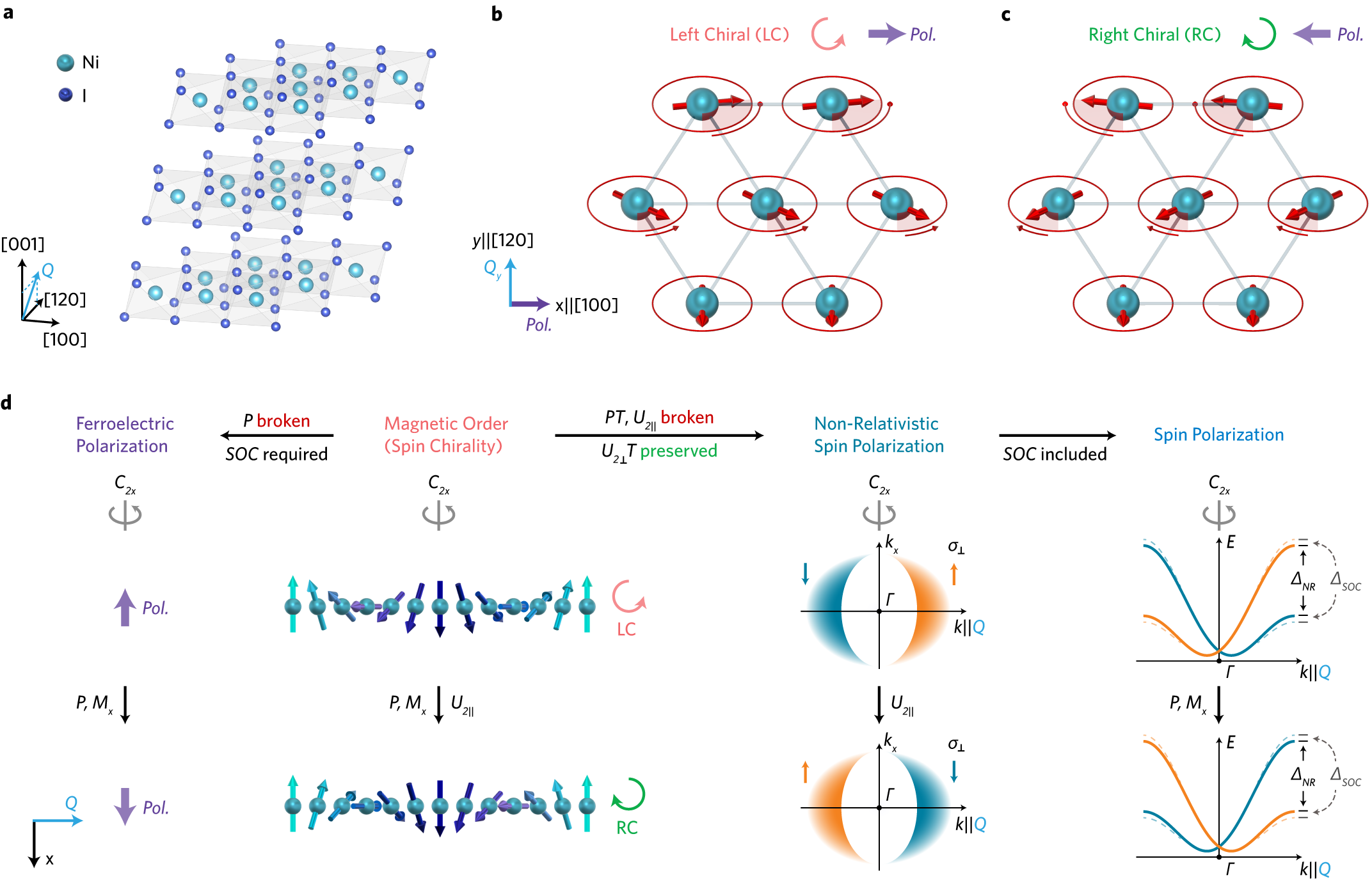}
\vspace{-10pt}
\renewcommand{\figurename}{Figure}
\linespread{1}
{
\caption{\textbf{Symmetry-protected magnetoelectric coupling in type-II multiferroic NiI$_2$}. \textbf{a}, Lattice structure of NiI$_2$ below $T_{N,2}$=59.5 K, with spin helix propagation vector $\mathbf{Q} = (0.138, 0, 1.457)$ reciprocal lattice units (r.l.u.), lying in the [120]/[001] plane. \textbf{b, c}, Magnetoelectric coupling between spin chirality and the spin-induced ferroelectric polarization ($\mathbf{Pol.}$). $\mathbf{Q}_{y}$ is the in-plane component of the spin helix propagation vector. The polar axis [100] is perpendicular to $\mathbf{Q}_{y}$. \textbf{d}, Symmetry-protected coupling between ferroelectric and (non-relativistic) spin polarization in NiI$_2$. On the one hand, the spin chirality of the chiral magnetic order breaks inversion symmetry, induces ferroelectric polarization and couples to it. On the other hand, the chiral magnetic order breaks $PT$ and $U_{2\parallel}$ symmetries, giving rise to odd-parity non-relativistic spin polarization $\sigma_{\perp}$, protected by $U_{2\perp}T$. $U_{2\parallel}$ and $U_{2\perp}$ represent two-fold spin rotation with axes parallel and perpendicular to the spin spiral plane. $\sigma_{\perp}$ represents spin perpendicular to the spin spiral plane. This non-relativistic effect promotes a significant spin splitting ($\Delta_{NR}$), much larger than that arising from SOC ($\Delta_{SOC}$). The symmetry operations that switch ferroelectric polarization will also switch (non-relativistic) spin polarization, allowing for $C_{2x}$ symmetry-protected electrical control of (non-relativistic) spin polarization. The solid lines represent the schematic of non-relativistic spin-polarized bands, and the dashed lines represent those that include spin-orbit coupling.}
\label{fig:fig1}
}
\vspace{-10pt}
\end{figure*}

The van der Waals (vdW) material NiI$_2$ has been identified as a spin spiral type-II multiferroic \cite{Kurumaji2013, song2022evidence, amini2024atomic}. NiI$_2$ adopts the rhombohedral $R\bar{3}m$ structure at room temperature, which comprises triangular lattices of Ni$^{2+}$ ions (with spin $S=1$) stacked along the $c$-axis (Fig. \ref{fig:fig1}a). The two-dimensional triangular lattice geometrically frustrates the intralayer magnetic exchange interactions between Ni spins \cite{Friedt1976, Amoroso2020, Botana2019, fumega2022microscopic}. This leads to a sequence of magnetic phase transitions, first to an antiferromagnetic (AFM) state at $T_{N,1} \simeq 75$ K, and then to a helimagnetic ground state below $T_{N,2} \simeq 59.5$ K (refer to Extended Data Figure 2 for structure and magnetometry characterization). The latter is characterized by an incommensurate, proper-screw spin helix with propagation vector $\mathbf{Q} \sim (0.138, 0, 1.457)$ reciprocal lattice units (r.l.u.). In terms of the real-space coordinates $x\parallel$[100], $y\parallel$[120], and $z\parallel$[001], Q lies on the $yz$ plane. The helix is tilted from the c-axis by $\sim 35$ degrees, with the spin spiral plane suggested to be nearly normal to $\mathbf{Q}$ \cite{Kuindersma1981}. As a result of SOC, which locks the spins to the lattice and implies that spin and spatial coordinates transform jointly, the helical magnetic order in NiI$_2$ breaks nearly all rotational and mirror symmetries of the host lattice, and only retains a single two-fold ($C_{2}$) axis perpendicular to the in-plane projection of the propagation vector $\mathbf{Q}$ \cite{arima2007ferroelectricity} (refer to Extended Data Figure 3 and Supplementary Material 1 for detailed symmetry analysis). Consequently, the helimagnetic transition is concomitant with the appearance of an in-plane ferroelectric polarization along the unique $C_{2}$ axis \cite{Kurumaji2013}, establishing NiI$_2$ as a spin spiral type-II multiferroic. Indeed, a pair of non-collinear spins generates, through a spin-current mechanism\cite{katsura2005spin}, an electric dipole with the general form $\mathbf{P}_{ij} = \mathbf{M}_{ij} \cdot (\vec{S}_i\times \vec{S}_j)$. Here, $i$ and $j$ are lattice sites, an arrow denotes an axial vector and boldface denotes a polar vector/tensor. $\mathbf{M}_{ij}$ is the antisymmetric part of a polar rank-3 tensor that describes the coupling between two non-collinear spins and electric dipole\cite{mike_PRL2011}. The sign of $\vec{S}_i\times \vec{S}_j$ denotes the (vector) spin chirality. In the aforementioned spin spiral, summing up over all bonds gives a total polarization that is proportional to the total spin chirality (see Supplementary Material 2 for a detailed symmetry analysis of $\mathbf{M}_{ij}$ and the resulting macroscopic polarization). It is important to emphasize that no symmetry operation exists which solely reverses either spin chirality or ferroelectric polarization while leaving the other unchanged, thereby establishing a symmetry-protected magnetoelectric coupling with record-strength \cite{gao2024giant} in NiI$_2$, as shown in Fig. \ref{fig:fig1}b, c. This strong coupling enables reliable electrical control of spin chirality through switching of the ferroelectric polarization, as observed in many other spin spiral multiferroics \cite{stein2017control, sagayama2010observation, yamasaki2007electric}.

A comprehensive analysis of the symmetry properties of a spin helix is essential for understanding the link between ferroelectric polarization and momentum-space spin polarization. The latter is quantified by the spin expectation value at momentum $\mathbf{k}$ and band $n$, $\left<\boldsymbol{\sigma} \right>=\frac{1}{2}\left<\psi_n(\mathbf{k})|\boldsymbol{\sigma}|\psi_n(\mathbf{k})\right>$, where $\psi_n(\mathbf{k})$ is the Bloch wave function and $\boldsymbol{\sigma}$ is the spin operator. Fig. \ref{fig:fig1}d illustrates how the symmetry properties of the spiral anchor the relationship between ferroelectric polarization, spin chirality, and spin polarization in NiI$_2$. As depicted in Fig. \ref{fig:fig1}d, the spiral in NiI$_2$ possesses a two-fold rotational symmetry axis along $x$ ($C_{2x}$), which specifies the unique axis of ferroelectric polarization (purple arrows) and the spin polarization nodal line ($k_y=k_z=0$). As a result, when inversion ($P$) or reflection with respect to the $yz$ plane ($M_x$) switches the spin chirality (orange and green arrows), the charge and spin polarizations switch as well, thereby establishing symmetry-protected coupling between these three quantities. Importantly, in order for these symmetry operations to act on both real space and spin space, one needs to invoke the existence of spin-orbit coupling (SOC). To establish a connection between these three quantities, one may be tempted to invoke a Rashba-like mechanism, by which a ferroelectric polarization generates an odd-parity spin polarization of the form $\left<\sigma_i(\mathbf{k}) \right>= -\left<\sigma_i(-\mathbf{k}) \right>$. While such a contribution indeed exists, the crucial point is that here the odd-parity spin polarization arises even in the absence of SOC \cite{Hellenes2024pwavemagnet,brekke2024minimal}.

To demonstrate this, we consider spin-group symmetry operations with rotations in spin-space independent from rotations in real space. Two-fold spin-space rotations about the $n$-axis are denoted as $U_{2n}$, distinct from magnetic-space rotations $C_{2n}$. As shown in the middle two panels of Fig. \ref{fig:fig1}d, the spin-helix breaks $PT$ symmetry but remains invariant under $U_{2\perp} T$, where $U_{2\perp}$ (two-fold spin rotation perpendicular to the spiral plane) compensates time-reversal $T$ \cite{Hellenes2024pwavemagnet}. In contrast, a two-fold spin rotation parallel to the spiral plane ($U_{2\parallel}$) reverses spin chirality (see Supplementary Material 1). These symmetries enforce an odd-parity spin polarization, $\left<\sigma_{\perp}(\mathbf{k}) \right>= -\left<\sigma_{\perp}(-\mathbf{k}) \right>$, where $\mathbf{k}\parallel \mathbf{Q}$, as illustrated in the non-relativistic spin polarization panel, leading to a non-relativistic spin splitting $\Delta_{NR}$. Additional symmetries may further constrain the odd-parity spin polarization. For instance, an even-period commensurate spiral retains $T\tau$ symmetry (where $\tau$ is a real-space translation), enforcing Kramer's degeneracy \cite{Hellenes2024pwavemagnet}. In the absence of SOC, while $T\tau$ symmetry enforces spin degeneracy in a collinear antiferromagent, it protects the odd-parity spin polarization in non-collinear non-centrosymmetric compensated magnets (see Supplementary Figure 3). Given that odd-parity spin polarization persists without SOC and is linked to spin chirality via spin-group symmetries, we now examine its behavior with SOC. As we will show in our DFT results, the magnitude of the spin splitting is only weakly affected by the SOC ($\Delta_{SOC}$). The primary role of SOC is to enable direct coupling between ferroelectric polarization and predominantly non-relativistic spin polarization, enabling electrical control and probing of spin polarization.


The odd-parity spin splitting in spiral magnets can also be understood via a simple model where itinerant electrons couple to winding magnetic moments through a Kondo-like interaction \cite{Loss2010,Choy2011,Martin2012,Flensberg2012,Nadj-Perge2013}. Recently linked to kinetomagnetism \cite{cheong2024kinetomagnetism, masuda2024room,chen2022designing, sandratskii1998noncollinear, Alex2024helimagneticsemimetal}, this model shows that in a one-dimensional spin helix, electronic bands experience an odd-parity Zeeman field \cite{Martin2012} $h_{\perp}(\mathbf{k}) = (\mathbf{k}\cdot\mathbf{Q})/2m$ polarized perpendicular to the spiral plane (see see Extended Data Figure 4a, b). This expression explicitly connects spin splitting and chirality ($h_{\perp} \sigma_\perp \propto k_\perp \sigma_\perp$) while revealing that spin splitting remains unaffected by the Kondo coupling, instead scaling with effective mass and wave-vector magnitude $Q$, suggesting a significant spin splitting for short-period spin helices (refer to Supplementary Material 3).

\section{Density Functional Theory calculations for the spin helix}

As discussed above, symmetry considerations indicate that NiI$_2$ exhibits odd-parity spin-splitting of predominantly non-relativistic character. While in experiments it is impossible to distinguish the two contributions (relativistic and non-relativistic), Density Functional Theory (DFT) offers an ideal framework to elucidate this issue. Therefore, we employ non-collinear DFT to analyze the momentum-space spin polarization in the helimagnetic state of NiI$_2$ (see Methods for computational details). Simulating the realistic incommensurate helimagnetism of bulk NiI$_2$ is computationally challenging, requiring large simulation cells and thus making the analysis of band structure and its spin polarization unnecessarily convoluted. Instead, we consider an isolated NiI$_2$ monolayer, leveraging its van der Waals nature and weak interlayer coupling (see Extended Data Figure 5). Within a single layer, the bulk helimagnetic state can be seen as a spiral propagating along the in-plane component of the propagation vector, with spins rotating in a plane that contains the two-fold C$_{2x}$ rotation axis perpendicular to the spiral wave vector. In our monolayer model, therefore, the spiral can be decomposed in a proper-screw helical part, with the spin spiral plane perpendicular to the in-plane propagation vector, and in a cycloidal component with spins rotating within the layer. Without loss of generality, we take the propagation vector parallel to the $y\parallel$ [120] direction and the spin spiral plane perpendicular to $z\parallel$ [001], i.e., a cycloidal-type spin spiral, as depicted in Extended Data Figure 6. The proper-screw component is equivalent from a symmetry perspective, as in both cases the ferroelectric polarization can only develop parallel to the two-fold rotation axis (see Supplementary Material 2 and 4), whereas the (non-relativistic) spin polarization is expected to develop perpendicular to the spin spiral plane.

\begin{figure*}[ht!]
\centering
\includegraphics[width=0.95\textwidth]{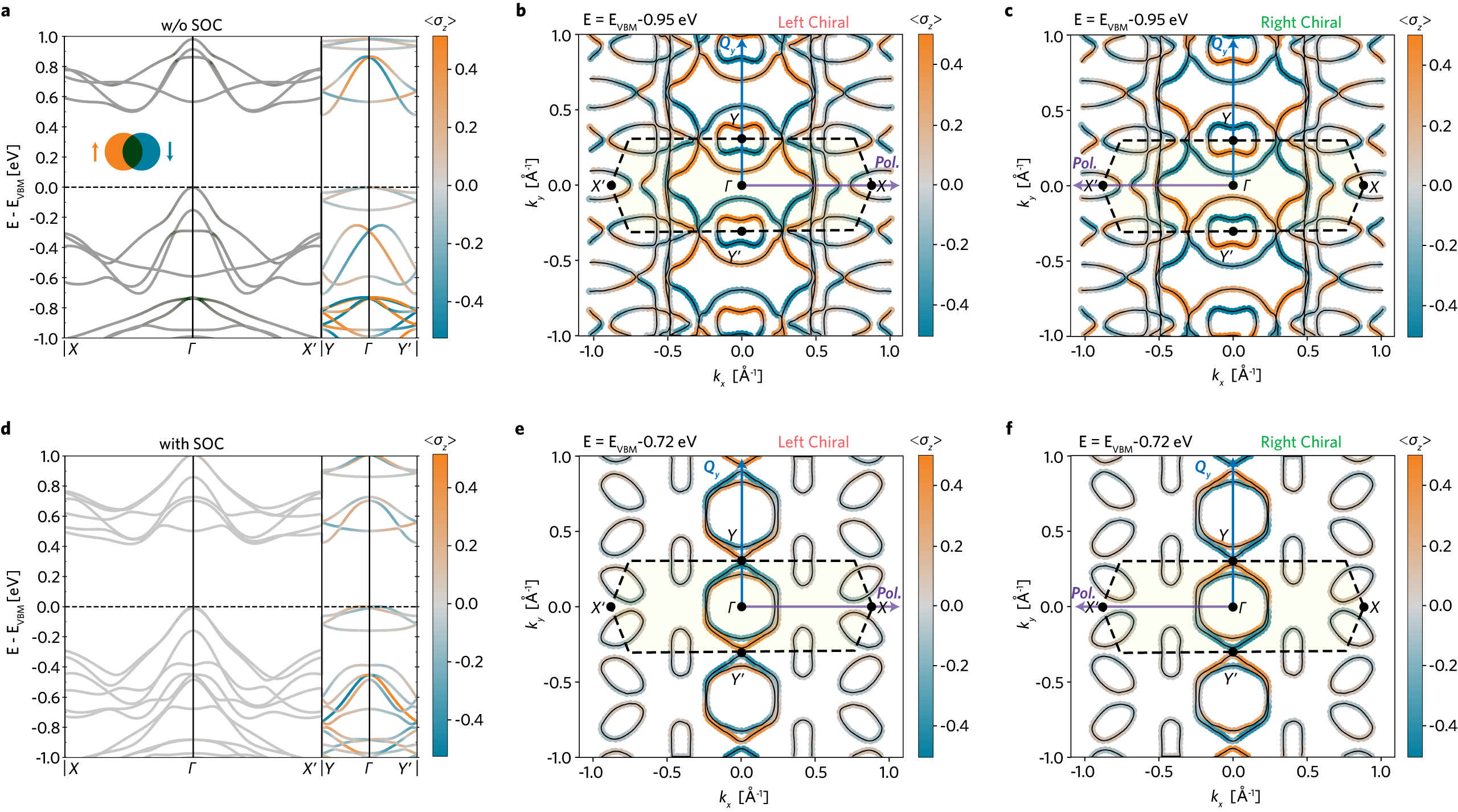}
\renewcommand{\figurename}{Figure}
\vspace{-10pt}
\linespread{1}
{
\caption{\textbf{DFT study of spin polarization in NiI$_2$}. Electronic bands of the period-3 spiral in NiI$_2$ superimposed with the momentum-dependent $\left<\sigma_z(\mathbf{k})\right>$ spin expectation values calculated along the band dispersion without SOC, \textbf{a}, and with SOC, \textbf{d}. The spin degenerate bands are coloured in gray. The ferroelectric polarization ($Pol.$) is along the $X-\Gamma-X'$ direction, while the spin helix propagates along $Y-\Gamma-Y'$. The $\left<\sigma_z(\mathbf{k})\right>$ expectation value vanishes along $X-\Gamma-X'$, but is non-zero and antisymmetric along $Y-\Gamma-Y'$, showing an odd-parity character. This $\left<\sigma_z(\mathbf{k})\right>$ spin polarization is protected by the $C_{2x}$ symmetry and persists even in the absence of SOC. $\left<\sigma_z(\mathbf{k})\right>$ spin polarization in the absence of SOC for left chiral, \textbf{b}, and right chiral, \textbf{c}, at constant energy cut $E=E_{VBM}-0.95 \, {\rm eV}$. The spin polarization is antisymmetric along $k_y$, and symmetric along $k_x$. The spin polarization reverses upon switching the spin chirality. $\left<\sigma_z(\mathbf{k})\right>$ spin polarization with SOC for left chiral, \textbf{e}, and right chiral, \textbf{f}, at constant energy cut $E=E_{VBM}-0.72 \, {\rm eV}$. The odd-parity character of $\left<\sigma_z(\mathbf{k})\right>$ spin polarization and its coupling to spin chirality are preserved regardless of SOC. $X$, $X'$, $Y$, $Y'$ are the high symmetry points at the boundary of the first Brillouin Zone (area within the dashed hexagon).}
\label{fig:fig2}
}
\vspace{-10pt}
\end{figure*}

Fig. \ref{fig:fig2}a shows the momentum-dependent expectation value $\left<\sigma_z(\mathbf{k})\right>$ for a period-3 left-handed helix without SOC. The expectation value vanishes along $X-\Gamma-X'$ but is non-zero and antisymmetric along $Y-\Gamma-Y'$ for both conduction and valence bands. Here, $X-\Gamma-X'$ is aligned to the ferroelectric polarization ($Pol.\parallel$ [100]) while $Y-\Gamma-Y'$ is along the in-plane spin helix propagation axis ($\mathbf{Q}_y\parallel$ [120]). Thus, the spin polarization is symmetric along the ferroelectric polarization direction but antisymmetric along $\mathbf{Q}_y$, protected by local $C_{2x}$ symmetry. In addition, both $\left<\sigma_x(\mathbf{k})\right>$ and $\left<\sigma_y(\mathbf{k})\right>$ vanish in the absence of SOC (see Extended Data Figure 5 and Supplementary Figure 2), supporting a spin polarization axis orthogonal to the spiral plane. This is further confirmed by rotating the spiral plane (Extended Data Figure 4c, d), where a proper-screw spiral in the $xz$ plane shows spin polarization axis parallel to $y$. Fig. \ref{fig:fig2}b present the non-relativistic spin polarization of the left-chiral helix at constant energy $E=E_{VBM}-0.95$ eV. Notably, as the spin chirality switches, the non-relativistic spin polarization also reverses, as shown in Fig. \ref{fig:fig2}c, supporting the coupling between the odd-parity non-relativistic spin polarization and spin chirality in NiI$_2$. The inclusion of SOC (Fig. \ref{fig:fig2}d) leads to additional spin splittings, but does not significantly change the magnitude at the $Y$ point, highlighting the dominance of the non-relativistic contribution. While details of the $\left<\sigma_z(\mathbf{k})\right>$ spin polarization are affected by SOC, thus depending on the spiral pitch and rotation plane, its odd-parity character and connection with spin chirality are consistently preserved, as seen in Fig. \ref{fig:fig2}e and f.

\section{Electrical switching of polarization in NiI$_2$}
\begin{figure*}[htb!]

\centering
\includegraphics[width=0.95\textwidth]{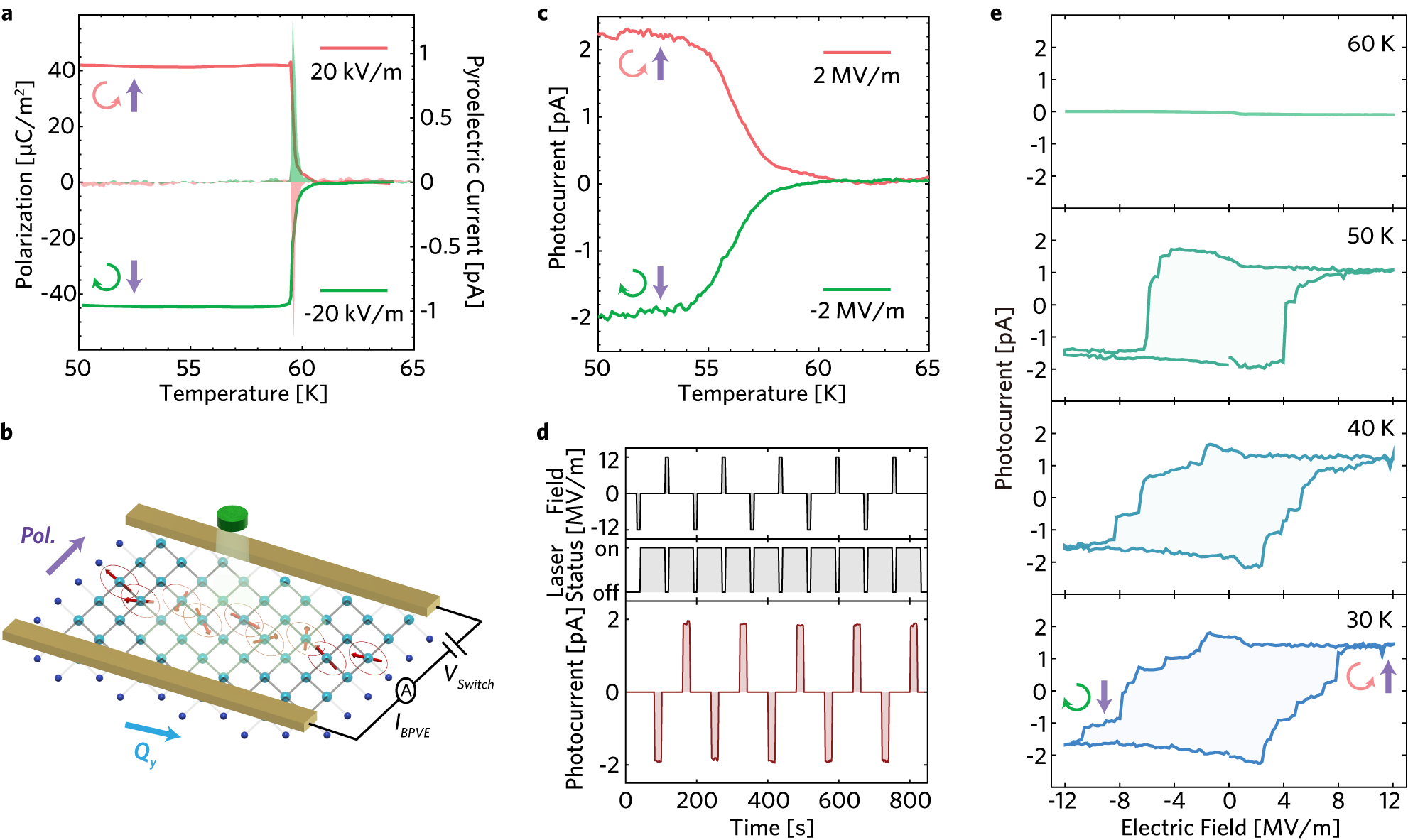}
\renewcommand{\figurename}{Figure}
\vspace{-10pt}
\linespread{1}
{
\caption{\textbf{Polarization switching of NiI$_2$}. \textbf{a}, Pyroelectric current measured in a bulk NiI$_2$ sample (shaded area). The electric polarization was extracted from time integration of the pyroelectric current (solid line). \textbf{b}, Schematics of zero-bias photocurrent measurement. The electrodes are placed along the ferroelectric polarization direction. Iodine atoms on top of Nickel are hidden for clarity. \textbf{c}, Temperature-dependent zero-bias photocurrent measurement in a 20 nm thick NiI$_2$ flake. \textbf{d}, Repeatable polarization switching by applying $\pm$ 12 MV/m pulsed electric fields at 30 K. The photocurrent was not collected at the dashed line region. \textbf{e}, Temperature dependent ferroelectric hysteresis loop measured by the zero-bias photocurrent method. In this figure, left and right spin chirality are represented as $\circlearrowleft$ and $\circlearrowright$; Photocurrent generated from the bulk photovoltaic effect is labelled as $I_{BPVE}$; The voltage required to switch the ferroelectric polarization is represented as $V_{Switch}$; Ferroelectric polarization along [100] is represented as purple arrow $\uparrow$ and $\downarrow$.}
\label{fig:fig3}
}
\vspace{-10pt}

\end{figure*}

Based on the above theoretical analysis, we expect that the spin chirality in NiI$_2$ can be reversed by switching the spin-induced ferroelectric polarization. Figure \ref{fig:fig3}a shows the ferroelectric polarization along the principal axis [100], measured using the pyroelectric current method in a bulk NiI$_2$ sample. An in-plane poling electric field of E = $\pm$20 kV/m was applied during cooling and removed prior to the current measurement in the warming process. A sharp peak in the pyroelectric current near 59.5 K reflects the depolarization at the multiferroic transition temperature (shaded region). With the poling electric field reversed, the pyroelectric current also changes sign, demonstrating the electrical switching of spin-induced ferroelectric order. A polarization of around 42 $\mu$C/m$^2$ was obtained upon time integration of the pyroelectric current (solid line). For micron scale NiI$_2$ samples, measurements of the electrical polarization become challenging, as the pyroelectric current scales with the sample dimension (see details in Extended Data Figure 7). Therefore, a local probe of polar order in a single spin helix domain region, which has typical lateral dimension of 10 $\mu$m \cite{song2022evidence}, is essential to investigate the unconventional odd-parity spin polarization, as different spin textures are realized in different domains, governed by local $C_2$ symmetries. 

In this study, we used zero-bias photocurrent as a sensitive local probe of polar order. A 20 nm thick NiI$_2$ was transferred with principal axis [100] aligned between two gold electrodes, as displayed in Fig. \ref{fig:fig3}b (see optical image in Extended Data Figure 7d). A poling electric field of E = $\pm$2 MV/m was applied to the NiI$_2$ flake in the cooling process, and the photocurrent was measured using 532 nm photoexcitation at zero bias in the warming process. The bulk photovoltaic effect in the multiferroic phase of NiI$_2$ is due to the built-in electrical polarization, producing a finite zero-bias photocurrent \cite{song2022evidence, grinberg2013perovskite} (refer to Extended Data Figure 7g-i for more details). Fig. \ref{fig:fig3}c shows the temperature dependence of zero-bias photocurrent for opposite field cooling. The intensity and sign of the photocurrent reflect the net ferroelectric polarization, which vanishes above the multiferroic transition temperature. This spin induced ferroelectric polarization can also be repeatably switched back and forth in the multiferroic phase, by applying $\pm$ 12 MV/m pulsed electric field at 30 K, as shown in Fig. \ref{fig:fig3}d. Complete ferroelectric hysteresis loops as a function of temperature were acquired by applying a sequence of pulsed electric fields, as shown in Fig. \ref{fig:fig3}e. The coercive and saturation fields are around 5 MV/m and 10 MV/m at 30 K, respectively. The kinks and sharp jumps in the hysteresis loop can be ascribed to domain redistribution after applying pulsed electric field \cite{matsubara2015magnetoelectric}, as zero-bias photocurrent only probes ferroelectric polarization in a micron-sized field of view. The coercive field is reduced as temperature is raised, and the loop closes above the transition temperature (60 K). The dependence of the zero-bias photocurrent signal on temperature and electric field confirms it is a sensitive local probe of polar order in spin spiral type-II multiferroic NiI$_2$.

\section{Odd-parity spin polarization in NiI$_2$}

The presence of odd-parity spin splitting driven by SOC in Rashba and Weyl systems has been previously investigated using the circular photogalvanic effect (CPGE) \cite{zhang2022room, niu2023tunable, wang2022circular, mciver2012control}. Because of the angular momentum of a circularly polarized pulse, the CPGE assesses optical transitions between electronic states with opposite spin angular momenta. To assess the odd-parity spin polarization in NiI$_2$, it is essential to perform the CPGE measurement within a single spin helix domain to avoid signal cancellation from multiple domains (refer to Extended Data Figure 8a for details of single domain selection procedure). NiI$_2$, with a triangular lattice, exhibits three pairs of equivalent spin helix domains \cite{Kurumaji2013, song2022evidence}. The propagation direction of the spin helix can be identified using linear dichroism, which distinguishes in-plane $\mathbf{Q}$ orientations through the associated breaking of three-fold rotational symmetry $C_{3z}$ \cite{song2022evidence}. Spatial mapping and angular dependence of linear dichroism in our flake reveal a single domain with propagation vector $\mathbf{Q}_{y}$ along [120] (Extended Data Figure 8b-d). Each propagation direction hosts left- and right-chiral domains with opposite ferroelectric polarization perpendicular to $\mathbf{Q}_{y}$, which can be controlled via electric field cooling (Extended Data Figure 8f). To probe odd-parity spin polarization using CPGE, the angular momentum of the incident light needs to have a component along the bands spin polarization axis, while the photocurrent detection needs to be along the direction in which the spin-texture is antisymmetric. In the case of NiI$_2$, we used 680 nm photoexcitation at normal incidence, since the axis perpendicular to the spin spiral plane has a large out-of-plane spin component, and the photocurrent detection is along the $\mathbf{Q}_{y}$ direction, which according to the symmetry analysis and DFT is the direction along which the $\left<\sigma_z(\mathbf{k})\right>$ spin polarization has odd-parity. 

\begin{figure*}[htb!]
\centering
\includegraphics[width=0.9\textwidth]{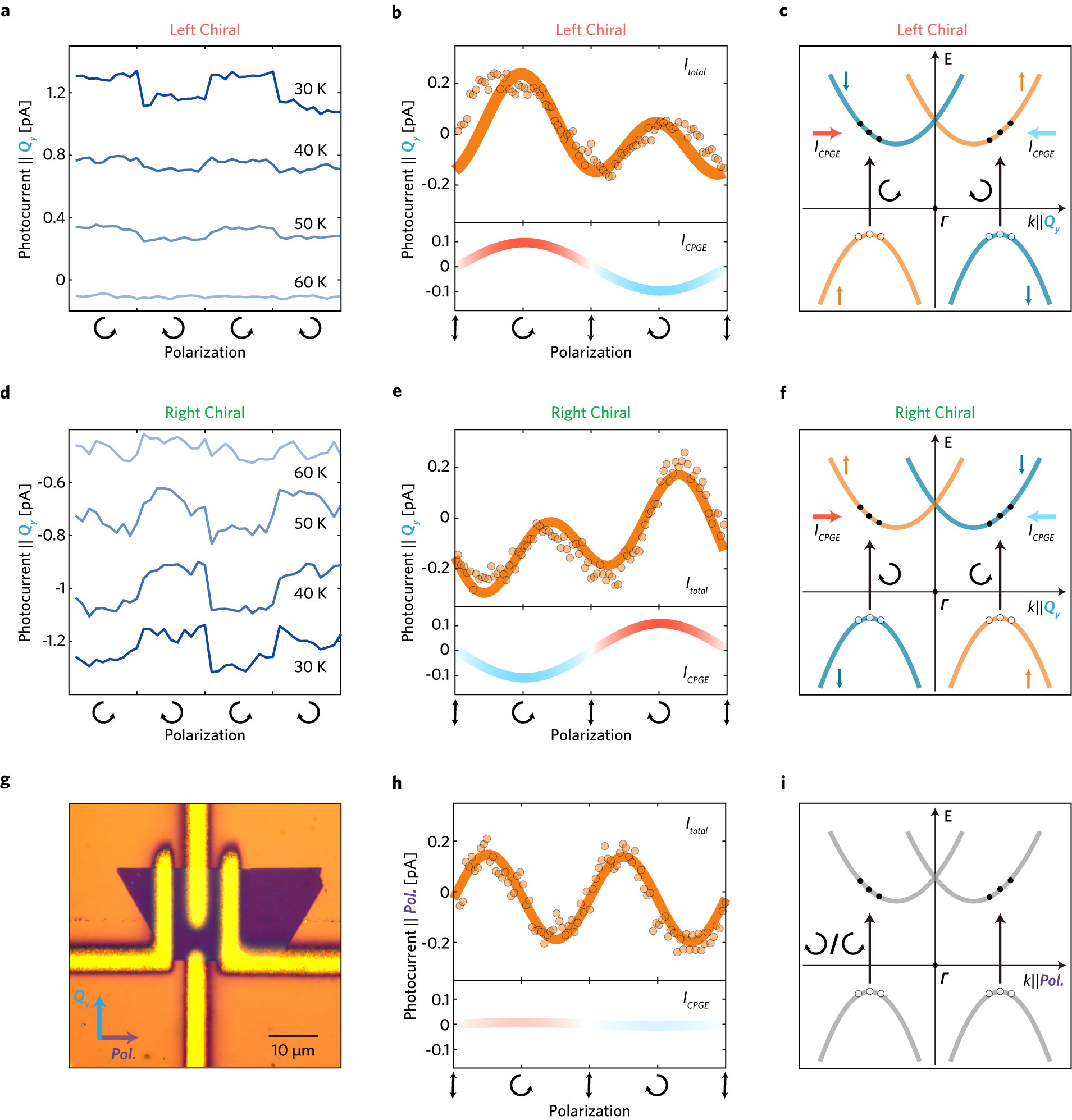}
\vspace{-10pt}
\renewcommand{\figurename}{Figure}
\linespread{1}
{
\caption{\textbf{Circular photogalvanic effect (CPGE) in NiI$_2$}. \textbf{a}, Temperature-dependent CPGE along $\mathbf{Q}_{y}$ for the left chiral spin helix. \textbf{b}, Polarization-dependent photocurrent along the $\mathbf{Q}_{y}$ direction shows a higher peak near LCP than RCP. \textbf{c}, Schematic spin polarization of the spin helix in NiI$_2$, highlighting its antisymmetric character along $\mathbf{Q}_{y}$ direction, which leads to a non-vanishing CPGE. \textbf{d}, Temperature-dependent CPGE along $\mathbf{Q}_{y}$ for the right chiral spin helix. \textbf{e}, Polarization-dependent photocurrent along the $\mathbf{Q}_{y}$ direction shows a higher peak near RCP than LCP. \textbf{f}, As the spin chirality switches, the spin polarization along $\mathbf{Q}_{y}$ also switches, leading to an opposite CPGE coefficient as compared with \textbf{c}. \textbf{g}, Optical image of the NiI$_2$ CPGE device. \textbf{h}, Polarization-dependent photocurrent along ferroelectric polarization direction does not display a discernible difference between LCP and RCP, implying that the CPGE is vanishingly small. \textbf{i}, The spin polarization of the spin helix in NiI$_2$ is symmetric along the $\mathbf{Pol.}$ direction, leading to a vanishing CPGE. The photocurrent generated from CPGE is labelled as $I_{CPGE}$; ferroelectric polarization is labelled as $\mathbf{Pol.}$; left and right circularly polarized light are represented as $\circlearrowleft$ (LCP) and $\circlearrowright$ (RCP).}
\label{fig:fig4}
}
\vspace{-10pt}
\end{figure*}


The CPGE device has two pairs of electrodes, as shown in Fig. \ref{fig:fig4}g. One pair is along the ferroelectric polarization direction [100], to switch and probe the ferroelectric polarization, and the other pair of electrodes is along the $\mathbf{Q}_{y}$ direction [120] to probe the odd-parity spin polarization. The incident light polarization is tuned by rotating an achromatic quarter-wave plate. The wavelength dependence of the CPGE has a maximum at 680 nm, near the charge transfer gap of NiI$_2$ (see Extended Data Figure 9a-d and 10i, j). Besides the CPGE current associated with circularly polarized light, there is also a spin-independent photocurrent when the incident light is linearly polarized. The total measured photocurrent can be expressed as $I_{total}(\phi)=C\sin(2\phi)+L_1\sin(4\phi)+L_2\cos(4\phi)+D$, where $\phi$ is the quarter wave plate angle, $C$ is the CPGE coefficient, $L_1$ and $L_2$ are the linear photogalvanic effect (LPGE) coefficients, and the constant $D$ contains other effects such as photovoltaic effect at the contacts, Dember effect and photon drag effects \cite{zhang2022room, niu2023tunable, wang2022circular, mciver2012control} (refer to Supplementary Material 5 for symmetry analysis of photogalvanic effects). The constant $D$ is subtracted in the angular dependence for clarity, and the fitting parameters are listed in the Methods section.

The NiI$_2$ device was first field-cooled to 30 K in the presence of an electric field of 1 MV/m along the ferroelectric polarization direction, which selects the left chiral spin helix ground state (Fig. \ref{fig:fig4}a-c). Fig. \ref{fig:fig4}a shows the temperature-dependent photocurrent along $\mathbf{Q}_{y}$, when the polarization of incident light is toggled between left (LCP) and right (RCP) circular polarization. In the multiferroic phase below 60 K, there is a clear jump of the photocurrent when switching between LCP and RCP. This jump vanishes at the transition temperature. The overall photocurrent amplitude also decreases as temperature is increased, which indicates the magnetic origin of the photocurrent. Fig. \ref{fig:fig4}b shows the angular dependence of the photocurrent along $\mathbf{Q}_{y}$, displaying a higher peak when illuminated with LCP light compared to RCP. This dependence of the photocurrent response on photon helicity is a consequence of the odd-parity spin polarization schematically shown in Fig. \ref{fig:fig4}c. Because LCP and RCP are related by a time-reversal operation, whenever LCP preferentially excites photocarriers with negative momentum along $\mathbf{Q}_{y}$, RCP must preferentially excite photocarriers with positive momentum, and vice versa. This effect leads to a non-zero CPGE. The NiI$_2$ device was then field cooled at -1 MV/m, corresponding to a right chiral spin helix (Fig. \ref{fig:fig4}d-f). Fig. \ref{fig:fig4}d shows that the relative amplitude of the photocurrent between LCP and RCP is reversed with respect to the case of the left chiral spin helix in Fig. \ref{fig:fig4}a, suggesting the switching of the odd-parity spin polarization. The angular dependence in Fig. \ref{fig:fig4}e also shows the switching of the CPGE coefficient when compared with Fig. \ref{fig:fig4}b, now with a higher peak at RCP than LCP, consistent with the band diagram in Fig. \ref{fig:fig4}f.

In contrast, the photocurrent along the ferroelectric polarization direction [100] barely shows any difference between LCP and RCP, as shown in Fig. \ref{fig:fig4}h. The analysis of the angular dependence reveals that the CPGE coefficient in this case is approximately two orders of magnitude lower than that along $\mathbf{Q}_{y}$. As we found in the symmetry analysis and confirmed by the DFT results, the spin polarization is expected to be symmetric along the $C_{2x}$ axis, thereby showing vanishing CPGE, as illustrated in Fig. \ref{fig:fig4}i. The observation of CPGE and its switching behavior is strong evidence of odd-parity spin polarization and its direct coupling to ferroelectric polarization in NiI$_2$.

\section{Discussion}

 
The observation of a robust odd-parity spin polarization in a type-II multiferroic with predominantly non-relativistic origin opens new opportunities for developing ultrafast, energy-efficient and high-endurance antiferromagnetic spintronic devices with substantial electrical readout capabilities. The electromagnon energy of the spin helix in NiI$_2$ is around 1 THz \cite{song2022evidence, Jae2023}, suggesting the potential for picosecond-scale switching of helimagnetic order \cite{gao2024giant}. While the spin-induced ferroelectric polarization is around four orders of magnitude lower than that of BaTiO$_3$ (a conventional ionic ferroelectric oxide), its coercive field is only half of that in BaTiO$_3$ \cite{martin2016thin}. Consequently, the switching energy can be expected to be up to five orders of magnitude lower. Furthermore, devices based on improper ferroelectrics of purely electronic origin are characterized by significantly higher endurance. Lastly, because of its dominant non-relativistic origin, the spin splitting of an odd-parity magnet can be as large as the eV scale, potentially resulting in enhanced electrical readout for spintronic devices\cite{qin2023room}. 

\section{Conclusion}

The reported results represent the first observation of an electrically-switchable unconventional odd-parity magnet. Group-theory analysis and DFT calculations confirm the odd-parity character of the spin polarization even in the absence of SOC. These findings open a new frontier to realize symmetry-protected voltage-based switching of non-relativistic spin polarization in a compensated magnet.

\section*{References}
\bibliography{references}

\newpage

\section*{Methods}

\subsection*{Growth and Characterization of NiI$_2$ Crystals}
NiI$_2$ thin flakes were grown on 300 nm SiO$_2$/Si via physical vapor deposition (PVD) in a horizontal single-zone furnace equipped with a 0.5 in. diameter quartz tube at ambient pressure\cite{liu2020vapor}. In a typical synthesis, 0.2 g of NiI$_2$ powder (99.5\%, anhydrous, Alfa Aesar) were positioned at the center of the furnace as the source material and the SiO$_2$/Si substrate was placed downstream at the maximum temperature gradient point. The furnace was purged by pumping the quartz tube below 0.5 Torr and then refilled with 99.99\% Ar gas two times. The furnace was heated to 430$^{\circ}C$ in 15 minutes and held at that temperature for 10 minutes. After the growth, the SiO$_2$/Si substrate was taken out immediately and stored inside a nitrogen-filled glove box (O$_2$ $<$ 0.2 ppm, H$_2$O $<$ 0.5 ppm). The sample thickness was determined by atomic force microscopy (AFMWorkshop HR) measurements performed inside a separate nitrogen-filled glovebox (O$_2$ $<$ 100 ppm, H$_2$O $<$ 1 ppm), using a silicon probe in tapping mode. The crystallographic orientation of NiI$_2$ flakes is determined by electron diffraction of transmission electron microscopy (FEI. Technai). Single crystal NiI$_2$ was grown by chemical vapor transport (CVT), from elemental precursors with molar ratio Ni:I=1:2, at a temperature gradient 650$^{\circ}$C to 570$^{\circ}$C. The magnetic susceptibility was measured during field cooling at 0.9 T applied out of plane, using a Magnetic Property Measurement System (MPMS-3, Quantum Design Inc.). X-ray diffraction of CVT grown crystals was performed in Bragg geometry using Cu K$_\alpha$ radiation (PANalytical).

\subsection*{Device fabrication}
The device for BPGE measurement was fabricated by depositing Ti(5 nm)/Au(30 nm) electrodes on SiO$_2$ (300 nm)/Si substrate using standard photolithography methods. A PVD grown NiI$_2$ flake was picked up and dropped down across the electrodes using a polymer-based transfer technique. To minimize the exposure to moisture, the polymer was dissolved in anhydrous chloroform inside the glovebox. The NiI$_2$ CPGE device was fabricated by transferring gold contacts on top of the sample. An ultrathin Si (50 $\mu$m thick) wafer with designed pattern was fabricated using deep-reactive-ion etching. Gold contacts (50 nm thick) without adhesion layer were deposited through the etched region onto the other SiO$_2$/Si handle wafer. These gold contacts without adhesion layer can be transferred using the standard polymer-based transfer technique.

\subsection*{Linear Dochroism Measurements}
A supercontinuum light source (SuperK, NKT Photonics) monochromatized to $\lambda = 633$ nm and a bandwidth of approximately $1$ nm was used as excitation for angular-dependent linear dichroism measurements, respectively. All measurements were performed at normal incidence in a closed-cycle optical cryostat (Opticool, Quantum Design). Linear dichroism measurements were performed with a photo-elastic modulator (PEM-200, Hinds Instruments) on the incident path of the optical setup. The beam incident on the PEM is prepared in linear polarization with an angle of $45^\circ$ with respect to the PEM fast axis and amplitude modulated with a mechanical chopper. The PEM retardance was set to 0.5$\lambda$ to modulate the incident polarization between $\pm 45^\circ$ linear polarization states. The light is then focused onto the sample using a 50x objective lens. The backscattered light is measured by an amplified photodiode (ThorLabs PDA100A2), whose output is connected to a lock-in amplifier (Stanford Instruments SR865A) referenced to the second harmonic of the fundamental PEM frequency $f = 50$ kHz. The total reflectance of the sample, used as a normalization, is monitored by a second lock-in amplifier referenced to the chopping frequency $f = 557$ Hz. This setup has a sensitivity down to 10 $\mu$rad.

To perform angular-dependent linear dichroism measurements, the angle of the perpendicular linear polarization states produced by the PEM is varied across the crystal using a zero-order half-wave plate placed just before the objective. In order to ensure the angular-dependence is recorded from a uniform, mono-domain region of the sample, linear dichroism microscopy images were first recorded at $T = 30$ K. The sample was held at this temperature for the duration of the angular-dependent measurements in order to maintain the same distribution of birefringent domains.

To perform linear dichroism imaging, the 50x objective is mounted on a closed-loop motorized stage. The polarization of incident light is aligned along [120] direction by rotating the half-wave plate. The scanning of linear dichroism was repeated at 30 K for two different cooling cycles to confirm the memory of the domain distribution.

Cross-polarized imaging was performed with a broadband visible LED light source, a standard CMOS-based monochrome camera and Glan-Thompson polarizers on both the input and output light paths in reflection geometry. A detuning of 0.5 (2.0) degrees from a cross-polarized configuration was used to maximize the contrast from birefringent domains.

\subsection*{Raman Spectroscopy Measurements}
Polarized Raman experiments were performed in a back-scattering geometry using a confocal microscope spectrometer (Horiba LabRAM HR Evolution) with a 50x objective lens and 532 nm laser excitation at a power of $40 \mu$W. Scattered light was dispersed by a 1800 lines/mm grating and detected with a liquid nitrogen cooled charge-coupled device (CCD) camera. The spectrometer integration time was 30 minutes, and each scan was taken twice and then averaged before analysis. An achromatic quarter-wave plate was placed in front of the objective with fast axis oriented at +/- 45 degrees with respect to the incident linear polarization for LCP/RCP circular incident polarization, respectively. For the reported circularly polarized spectra, no analyzing polarizer was used.

\subsection*{Second Harmonic Generation Measurements}
In second harmonic generation (SHG) measurements, an objective lens (Olympus LMPlanFL-N 50x) was used to focus an ultrashort laser beam onto the sample located in a cryostat (Janis ST-500). The laser fluence was set to 1 mJ/cm$^2$. Upon reflection, the second harmonic component of the beam radiated from sample was selected out by a dichroic mirror and a monochromator with 2 nm spectral resolution. The second harmonic photons were counted using a photomultiplier tube (Hamamatsu PMT) and a dual-channel gated photon-counter (Stanford Research SR400).

\subsection*{Transport Measurements}
The sub-pico ampere current was measured using electrometers (B2985B from Keysight) in a cryostat (Opticool from Quantum Design). The electrical connection uses a triaxial cable (three layers work as ground/shield/pin) from the electrometer to the cryostat. A coaxial cable (shield/pin) is used inside the cryostat, while the whole cryostat works as a ground layer. This triaxial connection is retained all the way from the electrometer to sample to reach femto-ampere noisefloor. A supercontinuum laser (SuperK, NKT Photonics) monochromatized by a prism and a bandwidth of approximately 1 nm was used for the photocurrent spectra measurement. The laser was focused using 20x objective to form a 5 $\mu$m diameter illumination on the NiI$_2$ sample for the zero-bias photocurrent measurement. To perform temperature-dependent zero-bias photocurrent measurement, the sample was first field-cooled from 80 K to 30 K in the dark. For the BPVE and CPGE measurements, the field cooling strength $\pm$ 2 MV/m and $\pm$ 1 MV/m was applied, respectively. Applying a higher electric field could show a leakage current at 80 K. Subsequently, the photocurrent was collected at zero bias while slowly warming at a rate of 0.5 K/min, to avoid thermocurrent effects. The temperature-dependent zero-bias photocurrent measurements in the CPGE devices (Extended Data Figure 8f and 10e) subtracted a constant background arising from the second pair of electrodes. For the ferroelectric hysteresis loop measurements, the laser shutter was closed while a 10 s pulsed electric field was applied, to avoid photocurrent overshot and minimize the risk of burning the device. At zero bias, the laser shutter was open to help stabilize the photocurrent and the zero-bias photocurrent was measured after a 60 s wait time. For pyroelectric current measurements, the edge of a $5 mm\times5 mm\times0.1 mm$ single crystal NiI$_2$ was cut to expose a fresh surface and the silver paste was applied to form electrical contact. The sample was first field cooled to the base temperature at a bias of $\pm$ 100 V, and then the current was measured in the dark at a warming speed of 3-4 K / min from 20 K to 80 K.

To perform CPGE measurement, circularly polarized light was prepared by placing a linear polarizer, followed by an achromatic quarter-wave plate (Thorlabs AQWP05M-600) before the objective. The laser was focused using a 50x objective to form a 2 $\mu$m diameter illumination on the NiI$_2$ sample. The circularly polarized photocurrent was collected by rotating the fast axis of the quarter-wave plate at 45$^{\circ}$ and -45$^{\circ}$ to the linear polarizer. The fitting parameters of the angular-dependent photocurrent are summarized in the table below. The linear polarization dependence $L_1$, $L_2$ and the constant shift $D$ could have an extrinsic origin, which is not studied in detail in this work.

\begin{table}[htb!]
\centering
\begin{tabular}{|>{\centering\arraybackslash}m{3cm} | >{\centering\arraybackslash}m{2cm}| >{\centering\arraybackslash}m{2cm}|>{\centering\arraybackslash}m{2cm}|>{\centering\arraybackslash}m{2cm}|} 
\hline
Parameters & $C$ & $L_1$ & $L_2$ & $D$ \\ 
\hline
Left Chiral & 0.0966 & 0.0147 & -0.1418 & 1.3995 \\ 
\hline
Right Chiral & -0.1085 & -0.1362 & -0.0745 & -1.4790 \\
\hline
Along Pol. & 0.0068 & 0.1684 & -0.0022 & 1.2540 \\
\hline

\end{tabular}
\end{table}

\subsection*{First-principles modelling of spin helices}
The electronic structure of spin helices in NiI$_2$ monolayer was calculated using the non-collinear DFT framework of the Vienna \textit{Ab-initio} simulation package ({\sc VASP}) based on pseudopotentials and plane waves\cite{Kresse1996Oct}. The structure of NiI$_2$ monolayer was modelled with a lattice parameter of $a = 3.97 \, \angstrom$, Ni-I bond length of $2.746 \, \angstrom$, and $25 \, \angstrom$ of separation along the $c$-axis ($z$-axis) to ensure the sufficient vacuum between periodic replicas. The effects of electronic exchange and correlation were described within the generalized gradient approximation using the Perdew–Burke–Ernzerhof (PBE) functional \cite{Perdew1996Oct}. The projector-augmented wave \cite{Blochl1994Dec} pseudopotentials with Ni $3p,3d,4s$ and I $5s,5p$ states in the valence were used and the Kohn-Sham wavefunctions were expanded on a plane-wave basis set with a cutoff of $500 \, {\rm eV}$. To keep the same level of accuracy for spin helices whose modeling requires employing different supercells (see Extended Data Figure 4-6), in each case the first Brillouin zone was sampled with a $k$-points mesh of density of $\simeq 0.009 \, \angstrom^{-2}$. Band structure and spin texture plots were produced using the {\sc Pyprocar} package\cite{Herath2020Jun}. Atomic Simulation Environment \cite{Larsen2017Jun} and {\sc XCrySden}\cite{Kokalj1999Jun} software were used for setting up DFT calculations and visualizing the spin helices.

\newpage

\subsection*{Acknowledgements}
\noindent This work was supported by the Department of Energy, Office of Science, Office of Basic Energy Sciences, under Award Number DE-SC0019126 (sample synthesis and device fabrication) and the National Science Foundation under Grant No. DMR-2405560 (photocurrent spectroscopy). The cryomagnet used in this work (OptiCool, Quantum Design) was acquired with support from the Air Force Office of Scientific Research under the Defense University Research Instrumentation Program (DURIP) Grant FA9550-22-1-0130. S.W.C. was supported by the W. M. Keck foundation grant to the Keck Center for Quantum Magnetism at Rutgers University. R.M.F. (theoretical model) was supported by the Air Force Office of Scientific Research under Award No. FA9550-21-1-0423. J.W.F.V. was supported by the National Science Foundation Award No. DMR-2144352. D.S.A. was supported by NSF Grant No. DMR-2410182 and Yale Mossman Postdoctoral Fellowship. S.S. acknowledges the financial support provided by the Ministry of Education, Science, and Technological Development of the Republic of Serbia. S.S., P.B., A.D., and S.P. acknowledge the CINECA award under the ISCRA initiative, for the availability of high-performance computing resources and support. B.I. , E.E and N.G acknowledge support from the US Department of Energy, Materials Science and Engineering Division, Office of Basic Energy Sciences (BES DMSE) for the SHG measurements.

\subsection*{Author contributions}
\noindent Q.S. and R.C. conceived the project. Q.S. synthesized the NiI$_2$ crystals, fabricated the devices, and performed optical and photocurrent measurements. C.A.O. provided support for optical measurements. S.S, P.B, A.D, and S.P. performed the DFT calculations and analysis. Q.S. and S.W.C. carried out the symmetry analysis. D.S.A., J.V., and R.M.F. carried out the group theory analysis. B.I. and E.E. performed SHG measurements supervised by N.G. Q.S., R.M.F., and R.C. wrote the paper with contributions from all coauthors. The authors declare no competing financial interests.

\subsection*{Competing interests}
The authors declare no competing interests.

\subsection*{Additional Information}
\noindent Correspondence and requests for materials should be addressed to Q.S. (qiansong@mit.edu), R.M.F. (rafaelf@illinois.edu), and R.C.(rcomin@mit.edu).

\subsection*{Data Availability}
The datasets generated during and/or analysed during the current study are available from the corresponding author on reasonable request.

\newpage
\setcounter{figure}{0}
\begin{figure*}[ht!]
    \centering
    \renewcommand{\figurename}{Extended Data Figure}
    \includegraphics[width=0.95\textwidth]{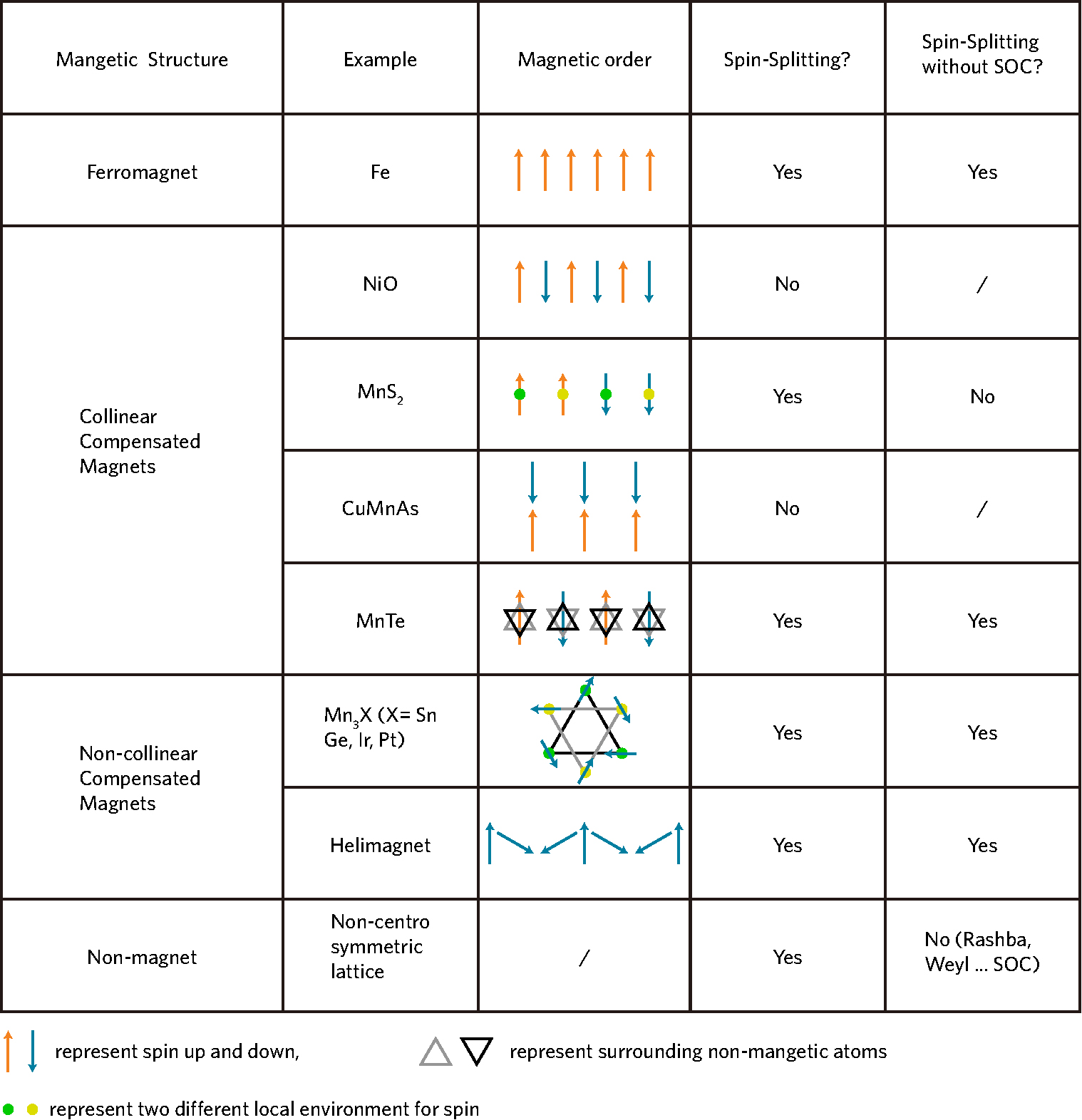}
    \linespread{1}
    {
    \caption{\textbf{Spin-splitting in different types of magnetically ordered states}. Depending on the spin-group symmetries that leave the magnetic configuration invariant, a spin splitting can emerge even in the absence of SOC.}
    \label{fig:figS1}
    }
\end{figure*}

\newpage

\begin{figure*}[ht!]
    \centering
    \renewcommand{\figurename}{Extended Data Figure}
    \includegraphics[width=0.95\textwidth]{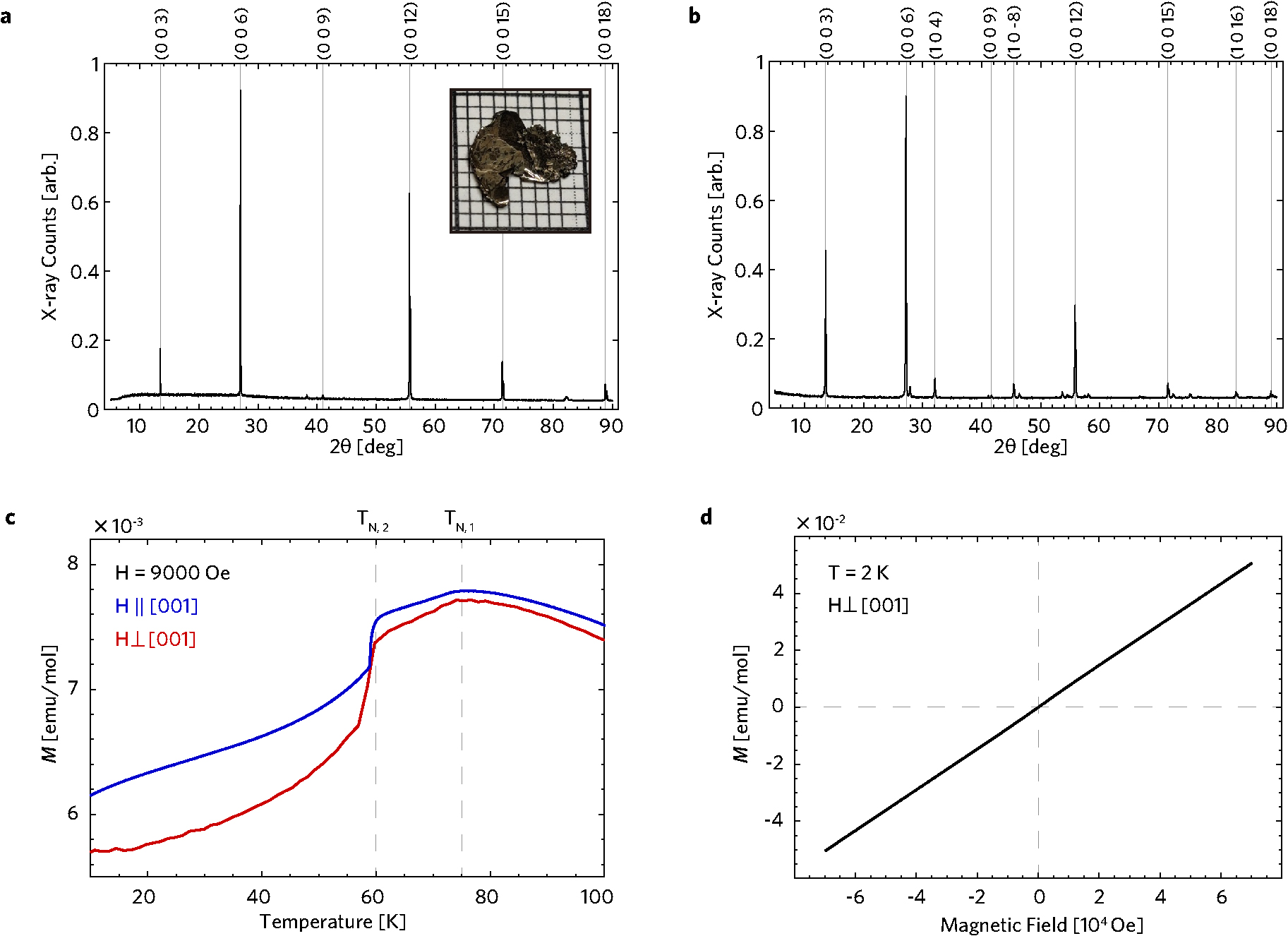}
    \linespread{1}
    {
    \caption{\textbf{Structural and magnetic characterization of NiI$_2$}. \textbf{a}, X-ray diffraction of a CVT-grown NiI$_2$ single crystal along [001] axis. Inset: optical image of NiI$_2$ single crystal with grid size 1 mm. \textbf{b}, X-ray powder diffraction of CVT-grown NiI$_2$ crystals. \textbf{c}, Temperature dependent magnetic susceptibility shows two magnetic transition, first to an antiferromagnetic state at $T_{N,1} = 75 K$, and then to a helimagnetic state below $T_{N,1} = 59.5 K$ with vanishing magnetization. \textbf{d}, Magnetic susceptibility at 2 K shows weak linear response to magnetic field up to 7 T. }
    \label{fig:figS2}
    }
\end{figure*}

\newpage

\begin{figure*}[ht!]
    \centering
    \renewcommand{\figurename}{Extended Data Figure}
    \vspace{-20pt}
    \includegraphics[width=0.95\textwidth]{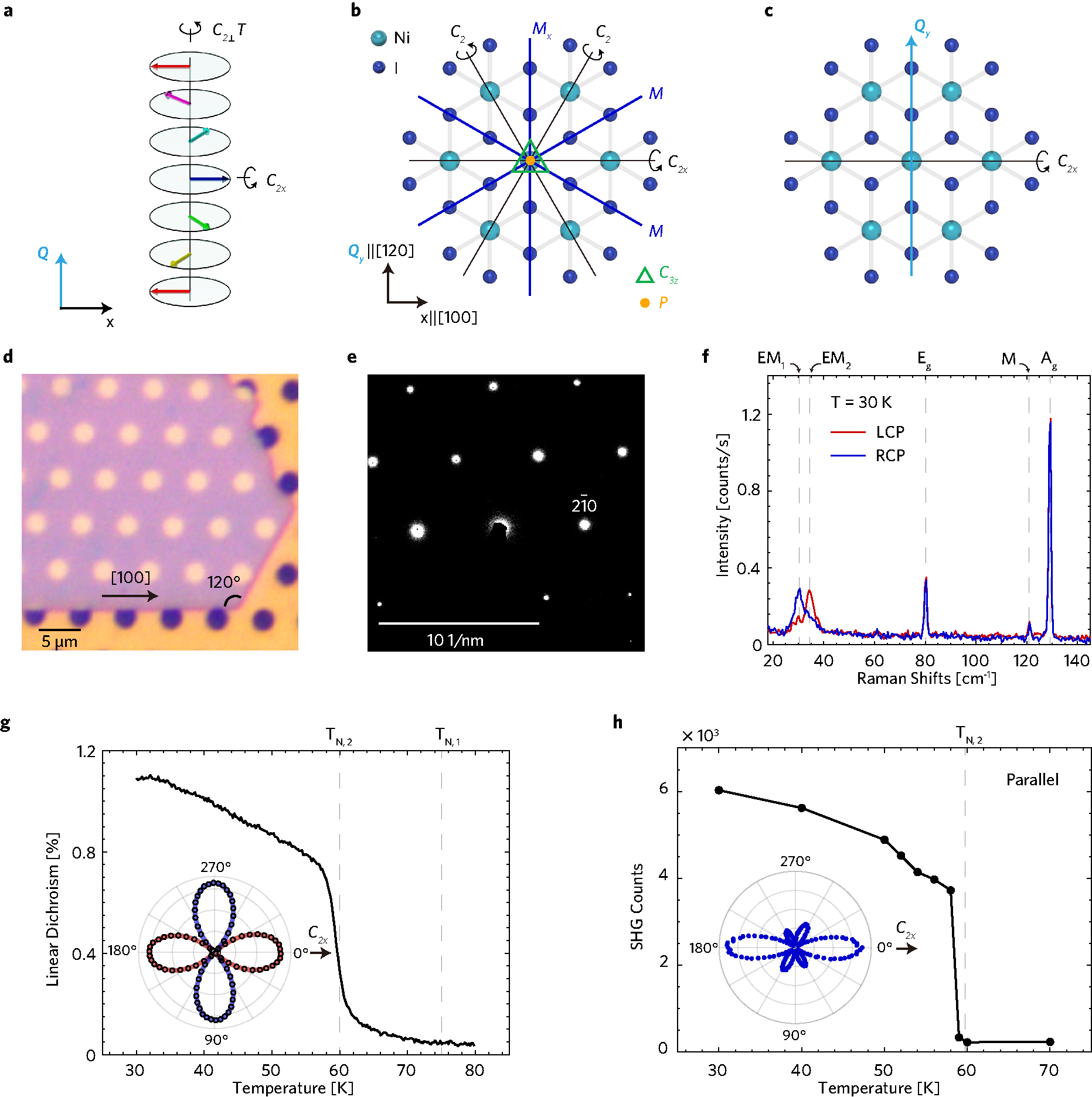}
    \linespread{1}
    {
    \caption{\textbf{Structural and Optical characterization of NiI$_2$}. \textbf{a}, Helical spin structure in a one-dimensional chain has multiple two-fold rotational symmetries C$_{2\perp}$T, and C$_{2\parallel}$ with rotation axes aligned with the local spin, e.g. C$_{2x}$. \textbf{b}, NiI$_2$ crystallizes in $R\bar{3}m$ structure, with three in-plane C$_2$ axes 60 degrees to each other. The C$_{2x}$ of the helix aligns with the C$_{2x}$ of lattice. \textbf{c}, Embedding the helical spin structure into $R\bar{3}m$ lattice with propagation vector $\mathbf{Q}$ lying in [120]/[001] plane results in a single two-fold rotational symmetry C$_{2x}$. \textbf{d}, Optical image of a 40 nm thick PVD grown NiI$_2$ flake transferred onto a Transmission Electron Microscopy (TEM) grid. \textbf{e}, TEM diffraction image is aligned with the optical image in \textbf{d}. The long edge of the PVD grown NiI$_2$ flake is confirmed to be the principle axis. \textbf{f}, Circularly polarized Raman spectroscopy at 30 K on a single-domain region of NiI$_2$ PVD grown flake shows circular dichroism in electromagnon modes (EM$_1$ and EM$_2$), but not in phonon modes (A$_g$ and E$_g$) or magnon mode (M). \textbf{g}, Temperature-dependent linear dichroism measurement on a single-domain region of NiI$_2$ PVD grown flake shows a sharp jump at T$_{N,2}$=59.5 K. The inset shows the angular-dependence of linear dichroism at 30 K. \textbf{h}, Below band gap ($\lambda = 991$ nm) temperature-dependent second harmonic generation (SHG) in parallel configuration on a single-domain region of NiI$_2$ PVD grown flake shows a sharp jump at T$_{N,2}$=59.5 K. The inset shows the angular-dependence of SHG at 30 K.}
    \label{fig:figS3}
    }
\end{figure*}

\newpage

\begin{figure*}[ht!]

\centering
\renewcommand{\figurename}{Extended Data Figure}
\includegraphics[width=0.85\textwidth]{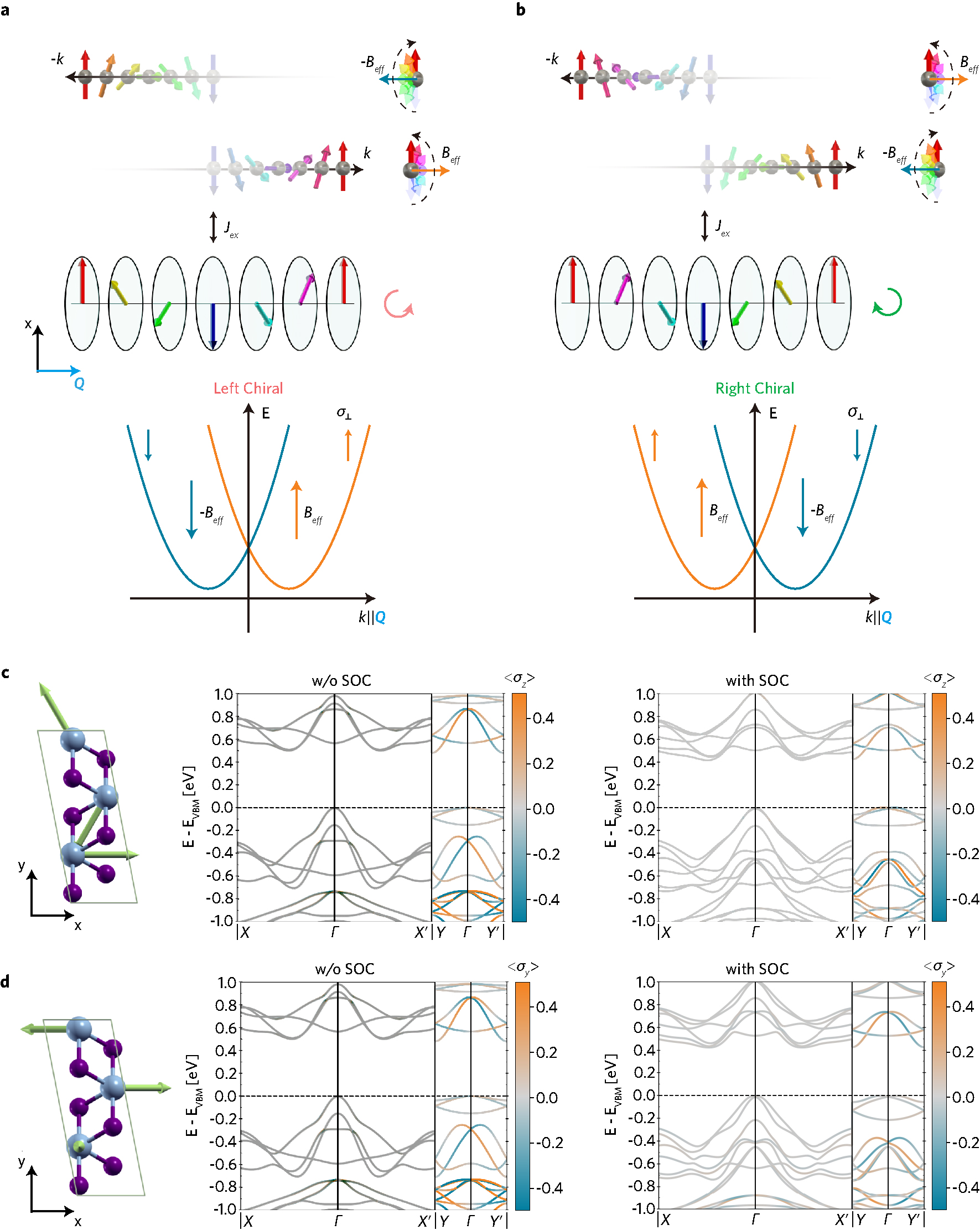}
\linespread{1}
{
\caption{\textbf{Schematics of spin-splitting in spiral magnets}. Itinerant electrons with opposite momenta $\pm k\parallel \mathbf{Q}$ experience opposite effective magnetic fields $\pm B_{eff}$ due to the Kondo interaction $J_{ex}$ with the localized magnetic moments of the spin helix. When the spin chirality switches, the sign of effective magnetic field also reverses, showing opposite momentum splitting of spin up and down channels in left chiral \textbf{a}, and right chiral \textbf{b}. See Supplementary Material Section 3 for further details on this model. \textbf{c}, The DFT model of cycloidal, \textbf{c}, and helical, \textbf{d}, component in NiI$_2$. The spin polarization is perpendicular to the spin spiral plane and antisymmetric along the propagation vector. In absence of SOC, the non-relativistic band structures are identical, with the spin polarization oriented perpendicular to the spin spiral plane. When SOC is included, relativistic effects on the band structure depend on the orientation of the spin-spiral plane. However, the odd-parity character of the non-relativistic spin polarization is preserved.}
\label{fig:figS4}
}
\end{figure*}



\newpage

\begin{figure*}[ht!]
    \centering
    \renewcommand{\figurename}{Extended Data Figure}
    \includegraphics[width=0.87\textwidth]{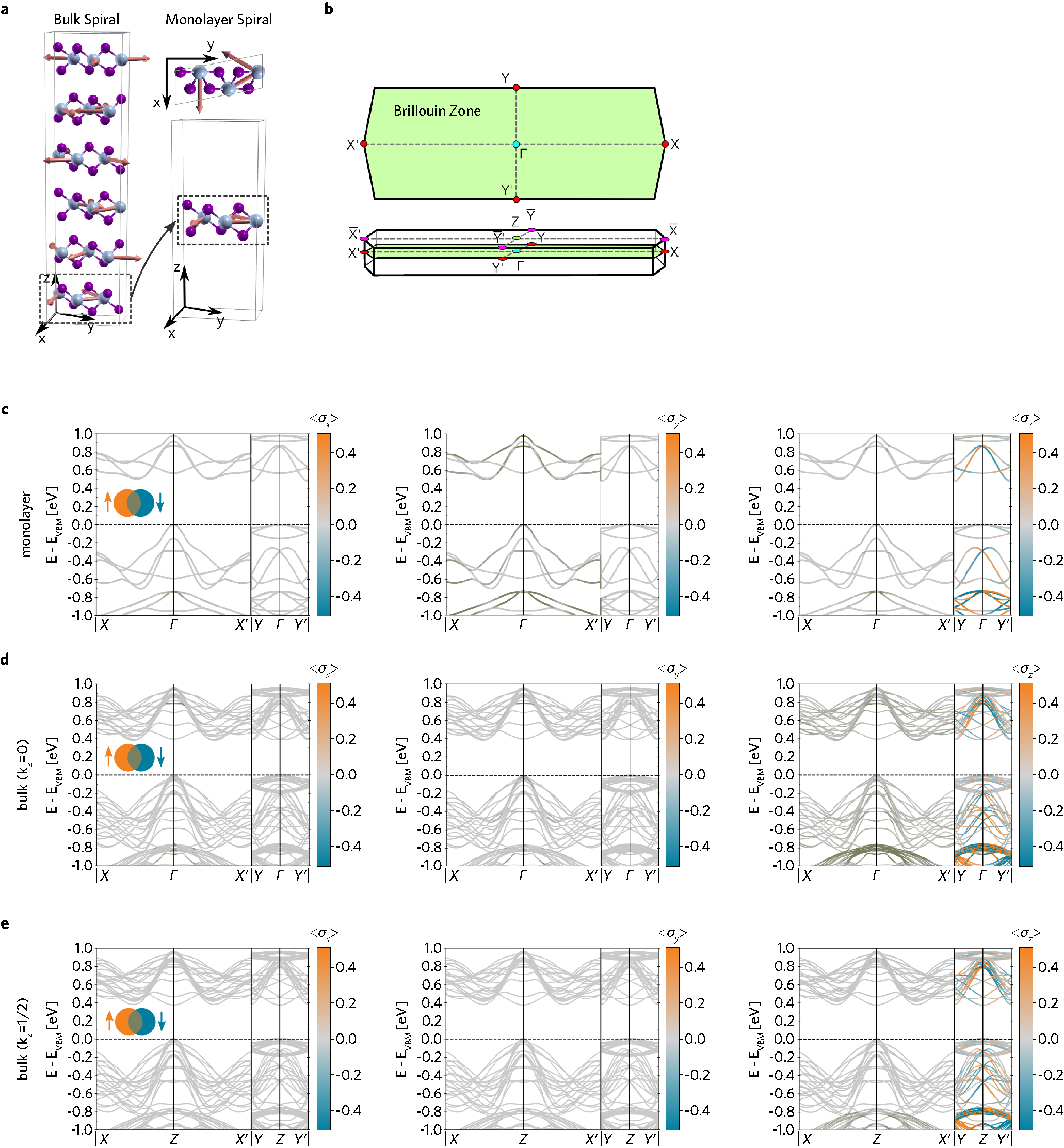}
    \linespread{1}
    {
    \caption{\textbf{Comparison of bulk and monolayer spirals}. \textbf{a}, Simulation cell for bulk NiI$_2$ with a spiral wave vector $Q=(1/3,0,3/2)$, compatible with the experimentally reported propagation vector but with a different pitch, highlighting the relationship with the adopted computational model for a single layer, alongside the corresponding Brillouin Zone \textbf{b}. Since the band structure is insensitive to the spiral plane in the absence of SOC, without loss of generality we have considered spins rotating in the $xy$ plane, realizing a pure cycloidal component in the monolayer case. The monolayer non-relativistic band structure \textbf{c}, is compared to the bulk one in the $k_z=0$, \textbf{d}, and $k_z=1/2$ planes, \textbf{e}, confirming the weak interlayer interactions and substantial vdW character of bulk NiI$_2$ as well as the same spin-polarization effect along the direction perpendicular to the spiral plane.}
    \label{fig:figS5}
    }
\end{figure*}

\newpage

\begin{figure*}[ht!]
    \centering
    \renewcommand{\figurename}{Extended Data Figure}
    \includegraphics[width=0.92\textwidth]{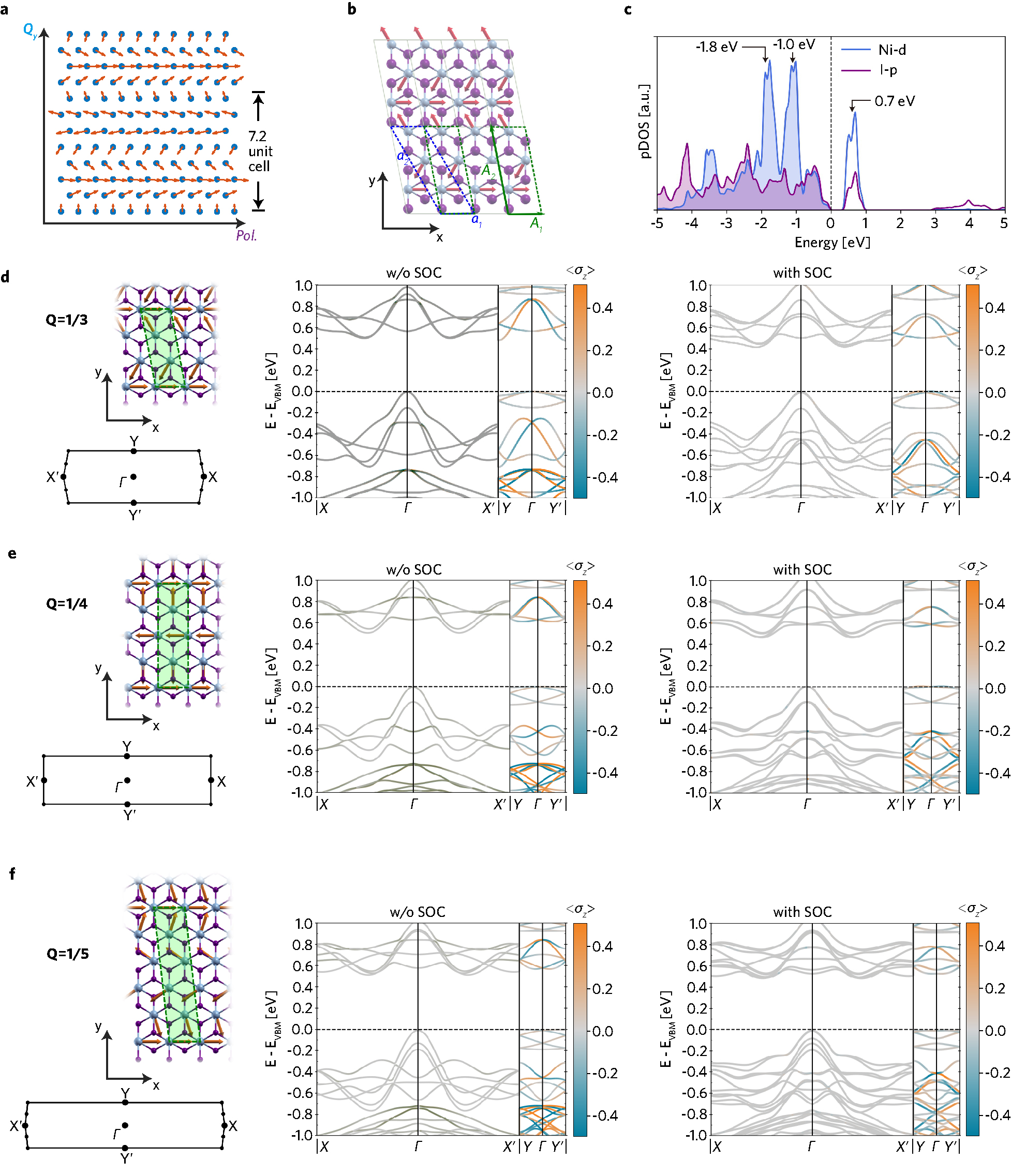}
    \linespread{1}
    {
    \caption{\textbf{DFT model of spin helix with propagation vector $Q=1/n$ (n is period)}. \textbf{a}, In-plane projection of spin helix in NiI$_2$ is a cycloid propagating along $\mathbf{Q}_{y}\parallel$ [120], inducing ferroelectric polarization along $\mathbf{Pol.}\parallel$ [100]. \textbf{b}, The DFT model of spin helix in NiI$_2$ was simplified as a cycloidal-type spin spiral propagating along $y\parallel$ [120], with spin spiral in $xy$ plane perpendicular to $z\parallel$ [001], inducing ferroelectric polarization along $x\parallel$ [100]. The period of the spin spiral for DFT calculation was three Ni spin (n=3), unless specified. The Wigner-Seitz cell (the green cell) was used for DFT calculation as its reciprocal cell represents the first Brillouin Zone. \textbf{c}, Projected Density of States of NiI$_2$ in the spin spiral phase. The most pronounced Ni-$3d$ character are at $E=E_{VBM}-1.8$, $-1.0$ and $+0.7$ eV. \textbf{d}, n=3. \textbf{e}, n=4. \textbf{f}, n=5. Magnetic supercells used to model the spin helices are presented together with their first Brillouin Zones. The $\left<\sigma_z\right>$ spin polarization is quantitatively influenced by periodicity and SOC, yet the odd-parity nature remains robust irrespective of them.}
    \label{fig:figS6}
    }
\end{figure*}

\newpage


\begin{figure*}[ht!]
    \centering
    \renewcommand{\figurename}{Extended Data Figure}
    \includegraphics[width=0.95\textwidth]{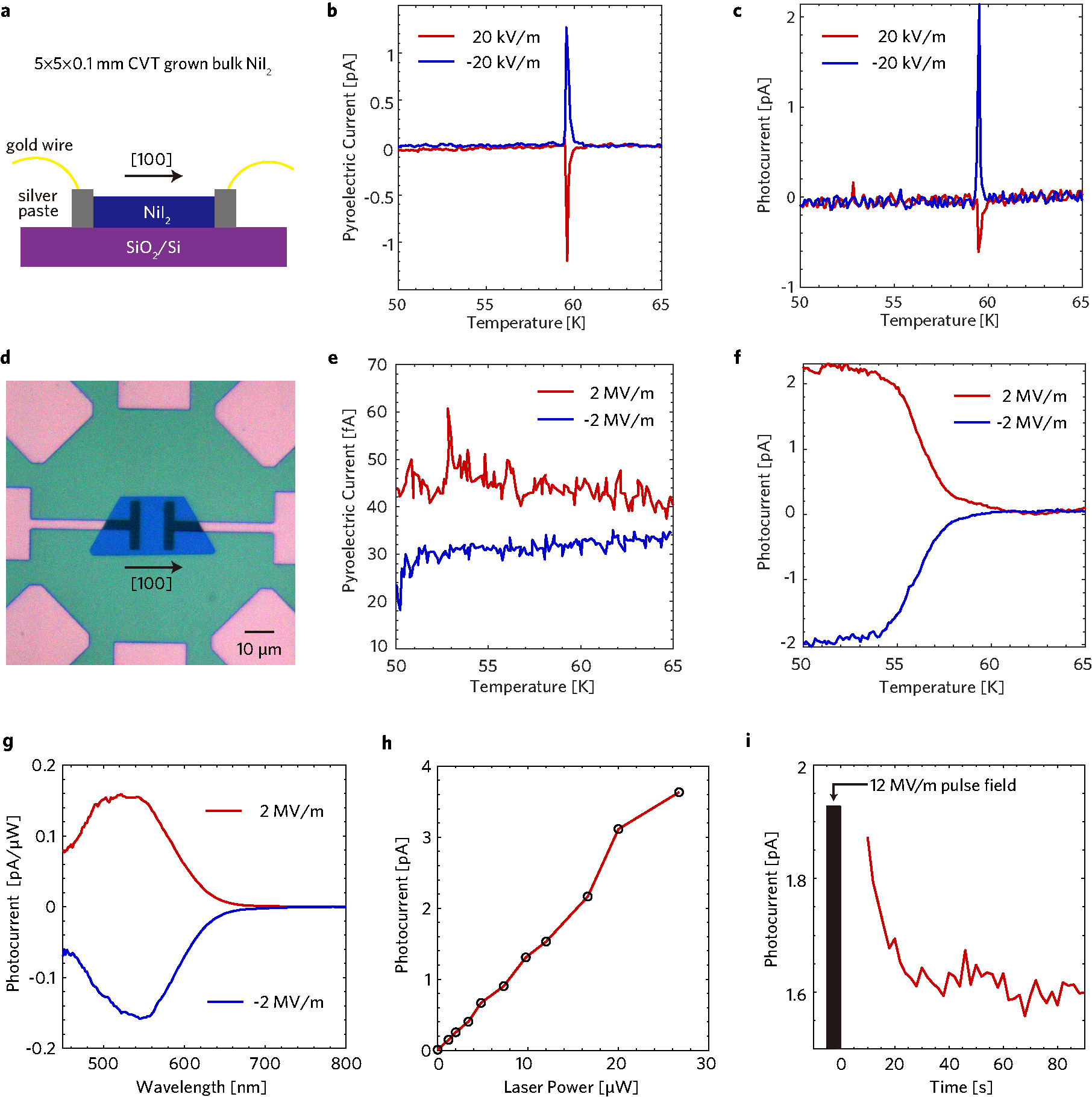}
    \linespread{1}
    {
    \caption{\textbf{Comparison of pyroelectric current and zero-bias photocurrent measurements}. \textbf{a}, Schematics of the CVT grown bulk NiI$_2$ device. \textbf{b}, Pyroelectric current of the device in \textbf{a}. \textbf{c}, Zero-bias photocurrent of the device in \textbf{a}. There is no clear photocurrent below the transition temperature, and the current is dominated by the pyroelectric current. The absolute value of pyroelectric current under illumination is not reliable. \textbf{d}, Optical image of a 20 nm thick PVD grown NiI$_2$ device. \textbf{e}, Pyroelectric current of the device in \textbf{d}. The pyroelectric current in a 20 nm thick flake is expected to be below atto-ampere, and there is no clear signature of multiferroic transition at femtoamp level in the temperature dependent measurement. \textbf{f}, Temperature dependent zero-bias photocurrent of the device in \textbf{d} shows clear transition near 59.5 K. \textbf{g}, Spectra of zero-bias photocurrent after opposite field cooling. \textbf{h}, Zero-bias photocurrent shows linear dependence with the laser power at 532 nm illumination. \textbf{i}, Stabilization time of zero-bias photocurrent after applying a 12 MV/m pulse electric field to switch the ferroelectric polarization. The current stabilized at around 40 s after the pulsed electric field. The laser was focused using 20x objective to form a 5 $\mu$m diameter illumination at the center of the NiI$_2$ flake for the bulk photovoltaic effect measurement.}
    \label{fig:figS7}
    }
\end{figure*}



\newpage

\begin{figure*}[ht!]
    \centering
    \renewcommand{\figurename}{Extended Data Figure}
    \includegraphics[width=0.95\textwidth]{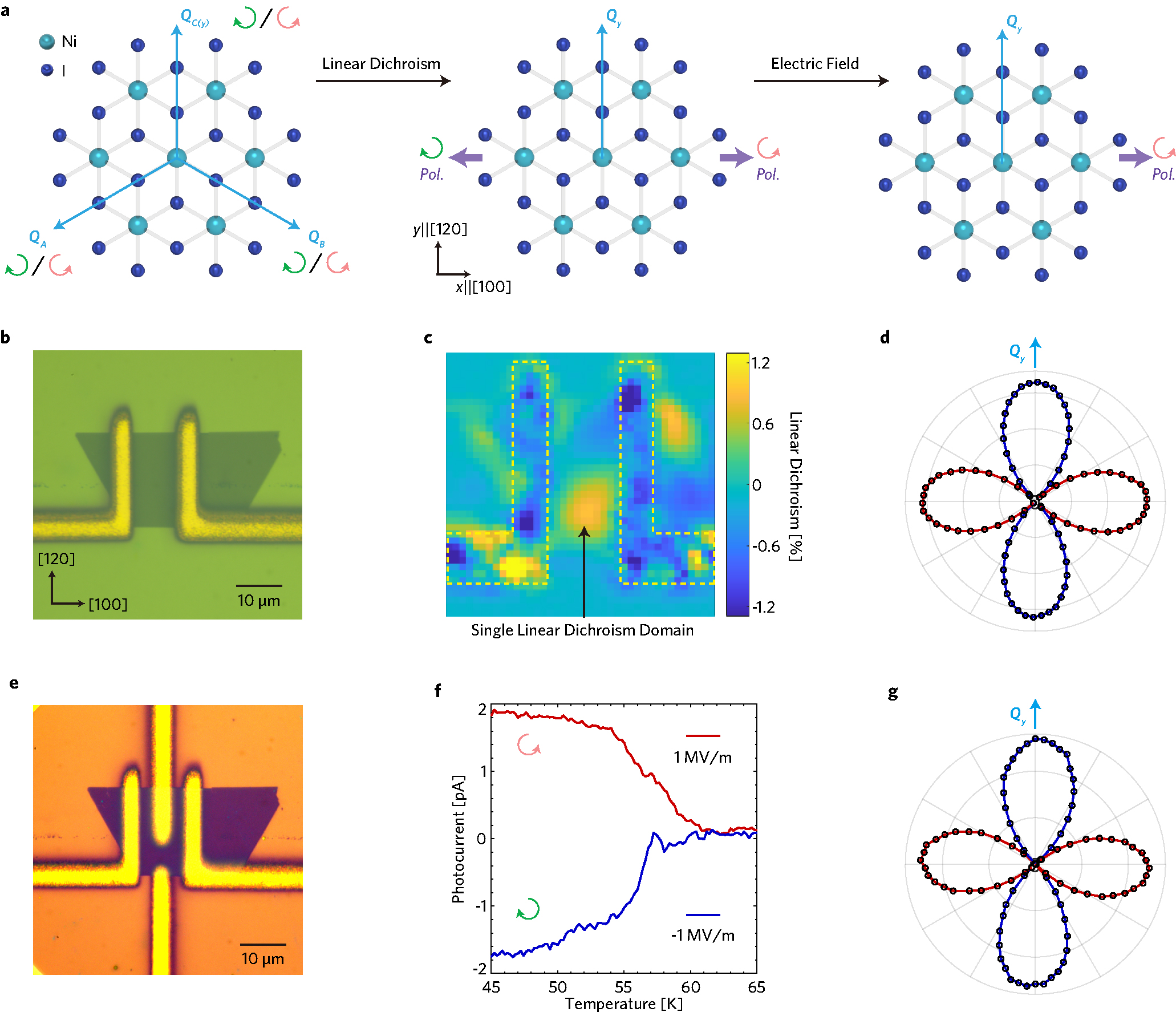}
    \linespread{1}
    {
    \caption{\textbf{Device fabrication for the CPGE measurement in NiI$_2$}. \textbf{a}, Schematics of a single domain selection procedure. In total, there are six multiferroic domains, with propagation vectors denoted as $\mathbf{Q}_{A}$, $\mathbf{Q}_{B}$ and $\mathbf{Q}_{C}$. Linear dichroism can distinguish the propagation direction, selecting a pair of chiral domains with opposite ferroelectric polarization ($Pol.$) perpendicular to the propagation vector ($\mathbf{Q}_{C}$ domain is denoted as $\mathbf{Q}_{y}$). Applying electric field further select a single spin helix domain.  \textbf{b}, Optical image of a NiI$_2$ flake with one pair of gold electrodes transferred along [100] direction. \textbf{c}, Linear dichroism scanning with light linearly polarized along [120] direction at 30 K. The yellow dashed lines mark the position of the gold electrodes. The state was prepared by 1 MV/m field cooling. \textbf{d}, Angular dependent linear dichroism at the single linear dichroism domain region marked in \textbf{c}. \textbf{e}, Another pair of gold electrodes were transferred along [120] direction at the single domian region. \textbf{f}, Temperature dependent zero-bias photocurrent along [100] direction shows the switching of ferroelectric polarization in the CPGE device after $\pm$1 MV/m field cooling. \textbf{g}, Angular dependent linear dichroism at the single domain region after transferring the gold electrodes at 30 K. Linear dichroism domain distributions are not sensitive to external electric field, but they have memory between different cool downs.}
    \label{fig:figS8}
    }
\end{figure*}

\newpage

\begin{figure*}[ht!]
    \centering
    \renewcommand{\figurename}{Extended Data Figure}
    \includegraphics[width=0.95\textwidth]{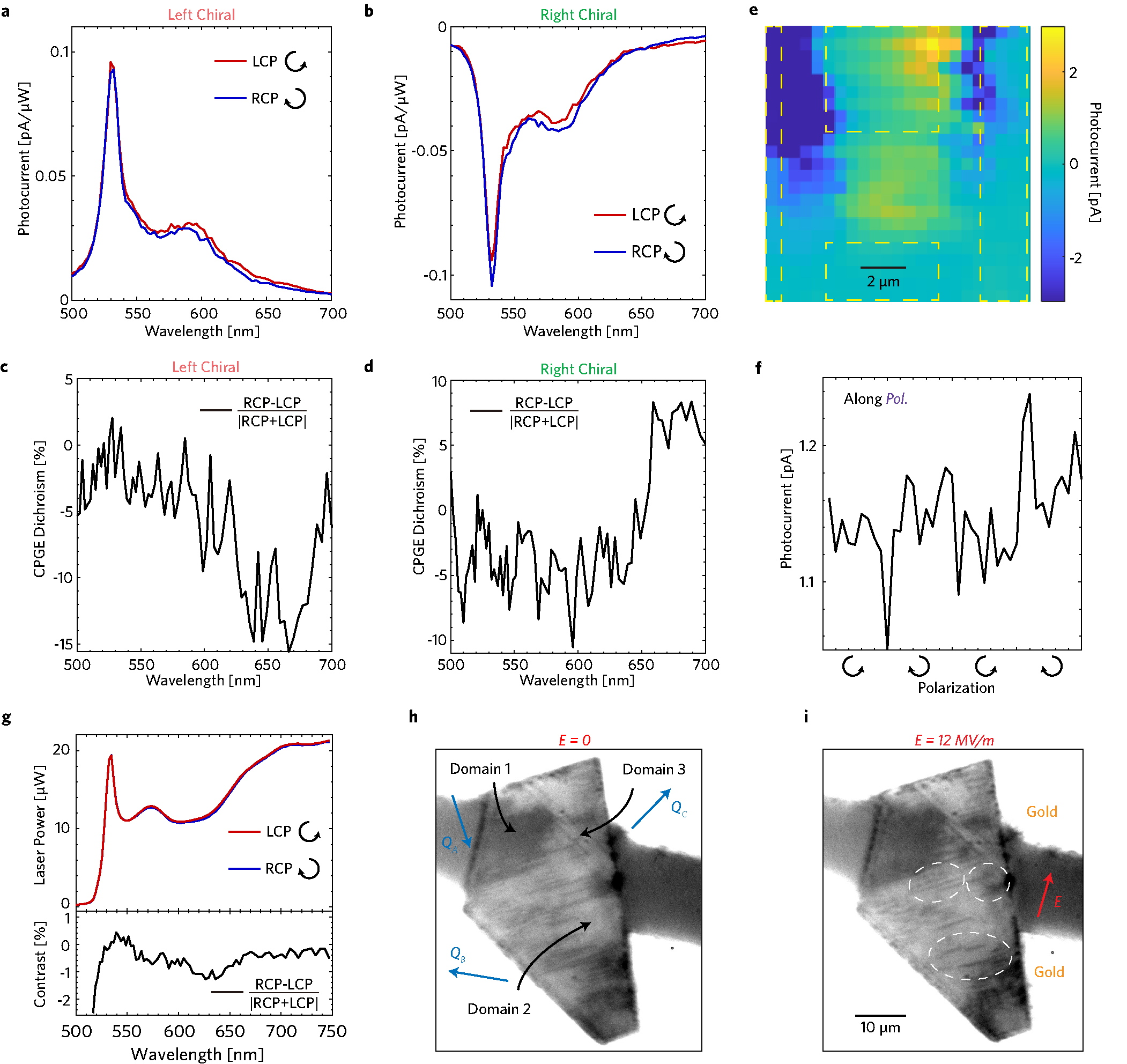}
    \linespread{1}
    {
    \caption{\textbf{CPGE spectra and imaging at 30 K}. Left and right circularly polarized photocurrent spectra along $\mathbf{Q}_{y}$ after 1 MV/m (\textbf{a, c}), and -1 MV/m (\textbf{b, d}), field cooling along ferroelectric polarization direction. The CPGE dichroism shows strongest contrast near 680 nm. \textbf{e}, Photocurrent along $\mathbf{Q}_{y}$ scanned at the sample region. The yellow squares mark the position of the gold electrodes. The sample region between the gold electrodes shows non-vanishing photocurrent. The laser was focused using a 50x objective and the spot size is around 2 $\mu$m. \textbf{f}, When the polarization of incident light was toggled between left and right circular polarization, the photocurrent along ferroelectric polarization did not show clear difference. \textbf{g},  Wavelength-dependent measurement of the laser power for both LCP and RCP at the sample position. Our optical setup introduces a 0.4\% artifact difference between LCP and RCP at 680 nm, which is substantially smaller than the CPGE observed in the experiment. \textbf{h}, Wide-field cross-polarized imaging for a 100 nm thick sample at zero-field cooling at 30 K. Some stripe-like domains are quite narrow with width less than 0.5 micron. \textbf{i}, Cross-polarized imaging of the same sample at 30 K with 12 MV/m field cooling. Overall, the domain distribution remains largely unchanged (memory effect), both between and on the electrodes. Some domains exhibit changes between temperature cycles, as indicated by the white circles. However, CPGE is still a robust probe of odd-parity spin polarization (refer to Supplementary Material 5).}
    \label{fig:figS9}
    }
\end{figure*}

\newpage

\begin{figure*}[ht!]
    \centering
    \renewcommand{\figurename}{Extended Data Figure}
    \includegraphics[width=0.9\textwidth]{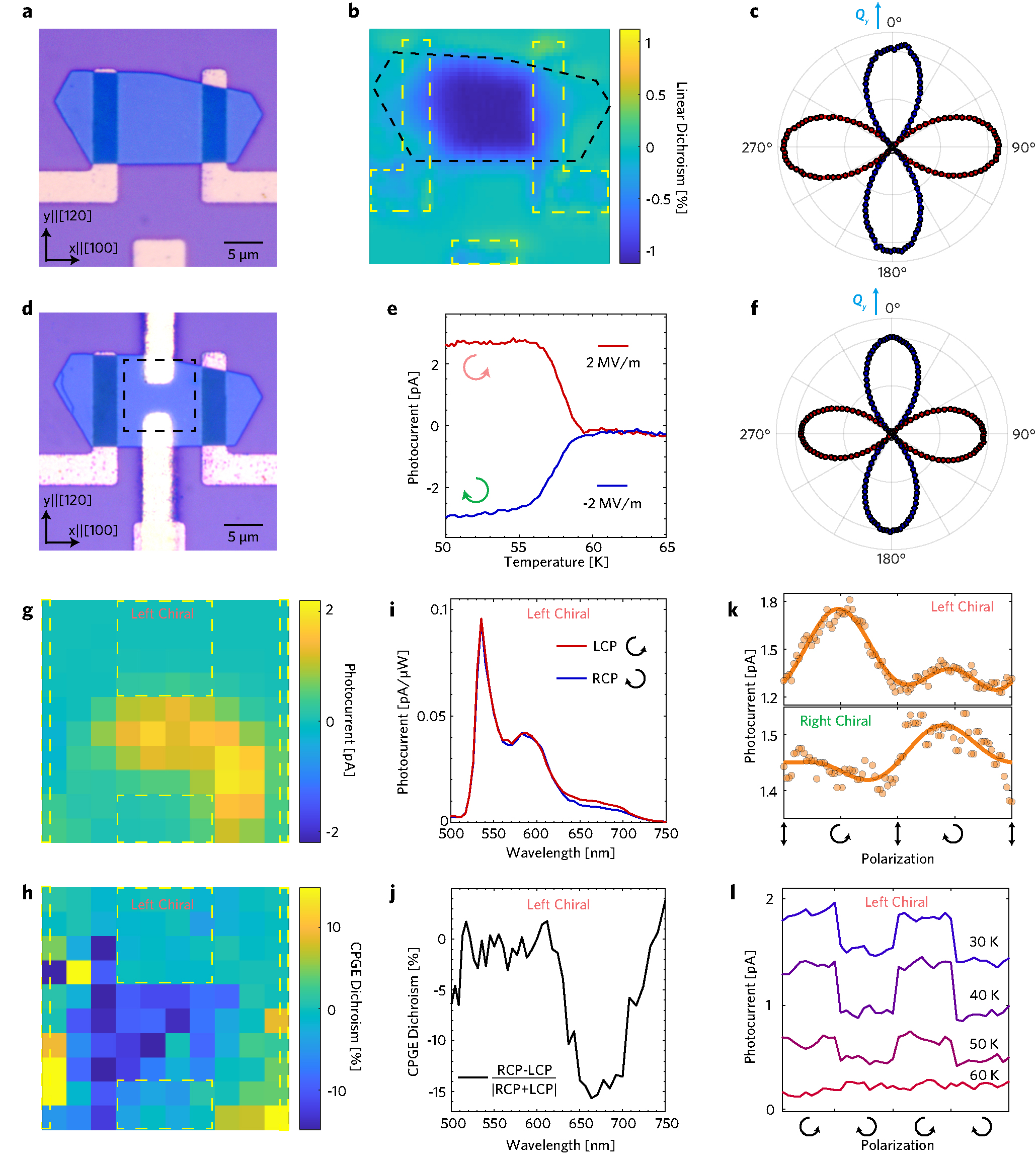}
    \linespread{1}
    {
    \caption{\textbf{Second CPGE device}. \textbf{a}, Optical image of the second CPGE device. \textbf{b}, Linear Dichroism (LD) imaging at 30 K shows a large single LD domain region between electrodes. \textbf{c}, Angular-dependent LD at 30 K shows that the single LD domain has propagation vector $\mathbf{Q}_y$ along [120] direction. \textbf{d}, Optical image after transferring one more pair of electrodes along $\mathbf{Q}_y$ direction. The square marks the region where the photocurrent scanning were done in \textbf{g} and \textbf{h}. \textbf{e}, Field cooling along $\mathbf{Pol.}$ shows switching of ferroelectric polarization. \textbf{f}, Angular-dependent LD at 30 K between top and bottom electrodes in \textbf{d} shows that the LD domain has memory. \textbf{g}, Photocurrent (Left Circurly Polarized 680 nm photoexcitation) along $\mathbf{Q}_y$ scanned at the sample region at 30 K. \textbf{h}, The CPGE scanning with 680 nm photoexcitation shows that the major contribution of CPGE is from the NiI$_2$ sample. The CPGE between left and bottom electrodes is due to the weak photocurrent at this region, which results in higher noise of CPGE. The laser was focused using a 50x objective forming a 2 $\mu$m spot size and each pixel of the mapping is 1 $\mu$m. \textbf{i}, Photocurrent spectra at 30 K shows a clear splitting between LCP and RCP at 630-720 nm, where the CPGE dichroism reaches -15\%, as shown in \textbf{j}. Angular dependence at 30 K, \textbf{k}, and temperature dependence, \textbf{l}, of CPGE with 680 nm photoexcitation. The change of CPGE coefficient for right chiral is likely due to domain redistribution in thermal cycles.}
    \label{fig:figS10}
    }
\end{figure*}

\end{document}